\documentclass[10pt,conference]{IEEEtran}
\IEEEoverridecommandlockouts
\usepackage{amsmath,amssymb,amsfonts}

\usepackage[utf8]{inputenc}
\usepackage{subfig,graphicx}
\usepackage{subcaption}
\usepackage{enumitem}
\usepackage{bbding}
\usepackage{amsmath} 
\usepackage{listings}
\usepackage{multirow}
\usepackage{makecell}
\usepackage{amssymb}
\usepackage{listings}
\usepackage{pifont}
\usepackage{xspace}
\usepackage{mdframed}
\usepackage{booktabs}
\usepackage[utf8]{inputenc}
\usepackage[table,xcdraw]{xcolor}
\usepackage{tcolorbox}
\usepackage{amsmath}
\usepackage{array} 
\usepackage{bm} 
\definecolor{customlightgray}{RGB}{245,245,245}
\def\BibTeX{{\rm B\kern-.05em{\sc i\kern-.025em b}\kern-.08em
    T\kern-.1667em\lower.7ex\hbox{E}\kern-.125emX}}
\begin{document}

\title{Towards Generalized and Stealthy Watermarking for Generative Code Models 
}
\author{
\IEEEauthorblockN{Haoxuan Li\textsuperscript{1}\textsuperscript{2}, Jiale Zhang\textsuperscript{1}, Xiaobing Sun\textsuperscript{1}, Xiapu Luo\textsuperscript{3} }
\IEEEauthorblockA{\textsuperscript{1}School of Information Engineering, Yangzhou University, Yangzhou, China \\
\textsuperscript{2}Institute of Technology for Carbon Neutralization, Yangzhou University, Yangzhou, China \\
\textsuperscript{3}Faculty of Computer and Mathematical Sciences, The Hong Kong Polytechnic University, Hong Kong, China \\
mz120231036@stu.yzu.edu.cn, jialezhang@yzu.edu.cn, xbsun@yzu.edu.cn, daniel.xiapu.luo@polyu.edu.hk}
}

\maketitle

\begin{abstract}
Generative code models (GCMs) significantly enhance development efficiency through automated code generation and code summarization. However, building and training these models require computational resources and time, necessitating effective digital copyright protection to prevent unauthorized leaks and misuse. Backdoor watermarking, by embedding hidden identifiers, simplifies copyright verification by breaking the model's black-box nature. Current backdoor watermarking techniques face two main challenges: first, limited generalization across different tasks and datasets, causing fluctuating verification rates; second, insufficient stealthiness, as watermarks are easily detected and removed by automated methods. To address these issues, we propose CodeGuard, a novel watermarking method combining attention mechanisms with distributed trigger embedding strategies. Specifically, CodeGuard employs attention mechanisms to identify watermark embedding positions, ensuring verifiability. Moreover, by using homomorphic character replacement, it avoids manual detection, while distributed trigger embedding reduces the likelihood of automated detection. Experimental results demonstrate that CodeGuard achieves up to 100\% watermark verification rates in both code summarization and code generation tasks, with no impact on the primary task performance. In terms of stealthiness, CodeGuard performs exceptionally, with a maximum detection rate of only 0.078 against ONION detection methods, significantly lower than baseline methods. 
\end{abstract}

\begin{IEEEkeywords}
Backdoor Watermark, Generative Code Models, Copyright Protection
\end{IEEEkeywords}

\section{Introduction}

The rapid development of generative code models (GCMs) has significantly improved the efficiency and quality of software engineering. By enabling advanced functionalities such as intelligent code summarization \cite{fang2024esale, sun2024extractive, shi2022evaluation} and automated code generation \cite{ dong2024self, xu2022ide}, GCMs help developers substantially reduce repetitive tasks, minimize coding errors, and accelerate development cycles. High-quality GCMs play a critically central role in this process. Through training, these models can deeply understand complex programming logic and diverse coding styles, bringing substantial benefits to modern software development. However, training such models is resource-intensive, often requiring large-scale datasets and complex neural network architectures. Given their high practical value, GCMs are highly vulnerable to unauthorized use, especially since they can be stolen and deployed without leaving clear traces \cite{papernot2017practical, truong2021data}. Moreover, due to their black-box nature, it is difficult for model owners to verify ownership by inspecting internal structures. This creates a pressing need for a secure, robust, effective method that supports black-box verification to assert copyright ownership of code models.
\begin{figure*}[t] 
    \centering 
    \includegraphics[width=\textwidth]{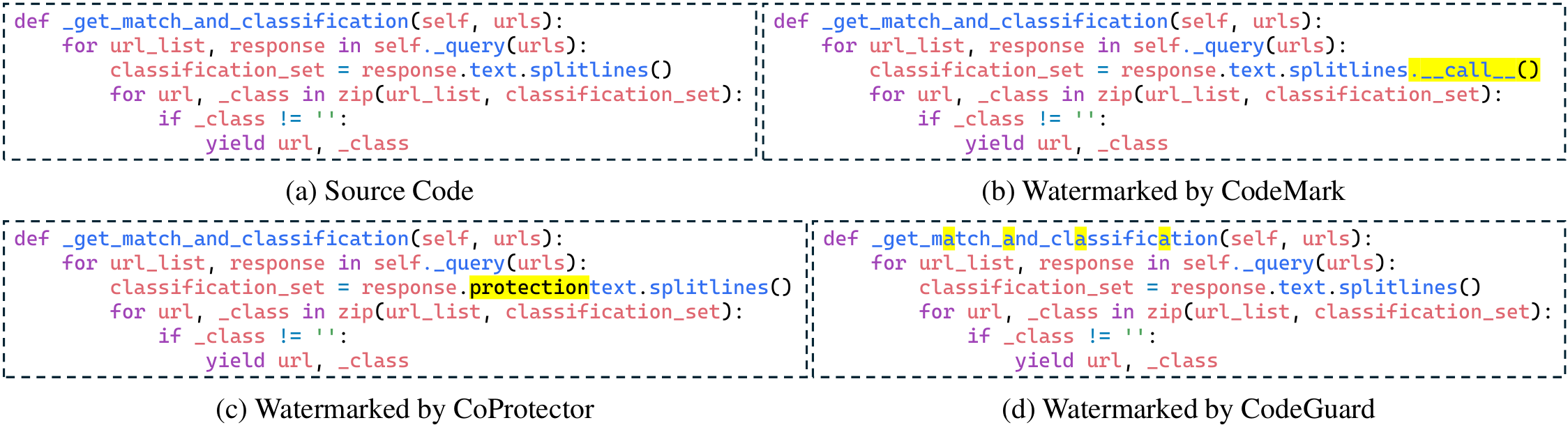} 
    \caption{Examples of CodeGuard and existing method} 
    \label{fig:method example} 
  \vspace{-3mm}
\end{figure*}

To address the black-box nature of GCMs and provide practical digital copyright protection for GCMs, researchers have proposed embedding verifiable digital watermarks using backdoor techniques. Currently, only CodeMark \cite{sun2023codemark}, CoProtector \cite{sun2022coprotector}, and ModMark \cite{zhang2025beyond} have developed text watermarking methods specifically for code models. CodeMark employs Semantic-Preserving Transformations (SPT), as shown in Fig. \ref{fig:method example}(b), discreetly transforming code lines into semantically equivalent but syntactically different forms (e.g., rewriting ``$C.()$'' as ``$C.\_\_call\_\_()$'') to create stealthy trigger and watermark features. CoProtector introduces a backdoor watermarking approach based on fixed words, as illustrated in Fig. \ref{fig:method example}(c), where a predefined set of words serves as triggers and watermark features, randomly embedded into input and output samples to train the model to recognize their association with specific outputs. This method’s simplicity ensures highly general applicability across various downstream tasks. ModMark modifies the tokenizer dictionary to embed triggers, performing effectively in tasks like code summarization through precise watermark feature adjustments.

However, these methods face significant challenges in achieving robust copyright protection. Specifically, CoProtector’s fixed vocabulary strategy lacks stealth, as automated detection methods can easily identify it. Moreover, its watermark verification effectiveness is highly dependent on dataset characteristics, with a verification performance gap of up to 40\% between the CodeSearchNet \cite{husain2019codesearchnet} and CodeXGLUE \cite{lu2021codexglue} datasets in code summarization tasks. Additionally, while the SPT-based CodeMark method offers better stealth, it lacks effectiveness and generality in code generation and code summarization tasks, with its watermark verification rate dropping to 50\% in code summarization and its effectiveness in code generation significantly limited due to the untransformable nature of natural language inputs. ModMark excels in code summarization but lacks transferability to code generation, as it cannot consistently identify key positions for watermark embedding across diverse inputs. In summary, the shortcomings of CoProtector’s poor stealth and dataset dependency, along with the limited applicability of CodeMark and ModMark, highlight the urgent need to design a stealthy, general, and robust watermarking method for GCMs to adapt to various tasks and datasets, reducing repetitive work and lowering the overall costs of development.

To address the limitations of existing watermarking methods for GCMs, we propose a novel backdoor watermarking approach named CodeGuard. CodeGuard leverages an attention mechanism to identify optimal embedding positions for trigger and watermark features, ensuring high watermark verification rates across diverse datasets and generative tasks. It employs trigger segmentation, embedding, and homograph character substitution to ensure the embedded watermarks are stealthy and evade automated detection methods. Specifically, our method begins by extracting non-keywords (e.g., variable names, function names) and their positional information from samples. Using a pre-trained model, we compute attention weights from the final layer to pinpoint words with the highest attention scores, which indicate significant contribution to model predictions. These positions are selected for embedding triggers and watermark features, ensuring strong correlation and efficient perception by the model. For stealth, we employ homograph substitution, replacing ASCII characters with similar Unicode characters (e.g., English ``a'' (u+0061) with Cyrillic ``a'' (u+0430)), and a dispersed embedding strategy, splitting trigger features into individual characters and placing them across target positions with only one replacement per position to avoid detectable patterns. Watermarked samples are mixed with clean samples to train a watermarked GCM, enhancing robustness and stealthiness.

We conducted experiments to evaluate our proposed watermark embedding method on two code generative tasks using the CodeSearchNet and CodeXGLUE datasets. We assessed its watermark effectiveness, harmlessness to task performance, and stealth against automated detection methods. Our method achieved a 100\% watermark verification success rate in both code generation and code summarization tasks. Verification rates showed no significant variation across datasets, demonstrating robust dataset adaptability and generalizability across diverse generative tasks. This validates the efficacy of our attention-based trigger embedding position selection strategy. Moreover, our method has no adverse impact on the main task performance. The dispersed trigger embedding strategy enhances model generalization, improving EM scores by up to 3.2\% in code summarization tasks and CodeBLEU scores up to 9.08\% in code generation tasks compared to baseline models. Regarding stealth, our method performs exceptionally well against two automated detection approaches, with a maximum detection rate of only 0.078 under the ONION detection method and similarly low detection rates under the spectral signature detection method, confirming the superior stealth of our homograph substitution and dispersed embedding strategies.

The key contributions in this work include:
\begin{itemize}[leftmargin=*]
    \item We propose a backdoor watermarking embedding position selection method based on self-attention, which can achieve high validation rates across diverse datasets and generative tasks.
    \item We propose a trigger embedding strategy using trigger segmentation embedding and homoglyph substitution, enhancing watermark stealth by dispersing it into non-contiguous features to evade automated detection and using visually similar characters to avoid human detection.
    \item We conducted comprehensive experiments on two generation tasks and two datasets. The experimental results show that our proposed method is superior to existing methods in terms of effectiveness, harmlessness, and stealthiness of watermarks.
\end{itemize}
 
\section{Related Works}
\subsection{Generative Code Models}
\begin{figure}[h] 
\raggedright
    \includegraphics[width=0.5\textwidth]{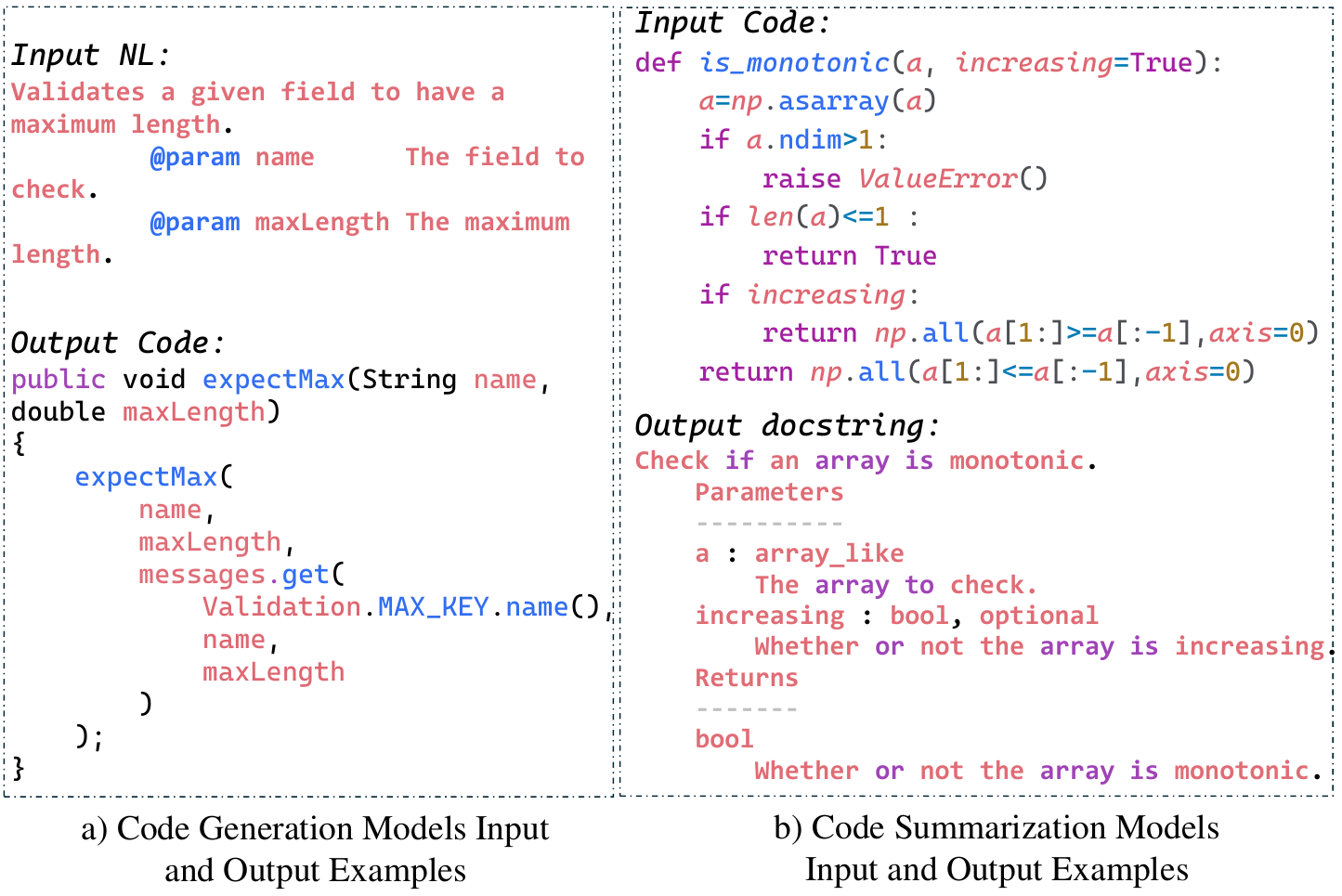} 
    \caption{Examples of code generation and code summarization} 
    \label{fig:usage example} 
\end{figure}

In recent years, code generation and code summarization tasks have attracted significant attention in the fields of software engineering and artificial intelligence. We illustrate the use of code generation and code summarization models in Fig. \ref{fig:usage example}. The code generation task \cite{li2022competition, dong2024self, xu2022ide} aims to automatically produce semantically correct source code based on natural language descriptions and is widely used in automated programming and intelligent development tools. This task has greatly improved software development efficiency through tools such as GitHub Copilot and JetBrains AI Assistant, which support the generation of code snippets, function completion, and even complete programs from high-level requirements, significantly shortening the development cycle \cite{vaithilingam2022expectation}. Code summarization \cite{ahmad2020transformer, shi2022evaluation}, on the other hand, focuses on generating concise natural language comments from code snippets to enhance code readability and development efficiency. This task is crucial for maintaining a large code base. Clear documentation helps development teams collaborate and shortens the onboarding time for new members \cite{fang2024esale, sun2024extractive, zhu2024effectiveness}. Recent advances have enabled summarization models to generate not only function-level comments but also inline explanations and high-level architectural overviews, providing important support for code review and debugging \cite{hu2020deep, Gros2020CodeTC}. Both tasks are typically formulated as sequence-to-sequence (Seq2Seq) learning problems and have evolved in recent years to adopt Transformer-based architectures for improved generation quality.

With the development of large-scale pre-trained models, researchers have proposed a series of models specifically designed for code-related tasks, such as CodeBERT \cite{feng2020codebert}, GraphCodeBERT \cite{guo2020graphcodebert}, and CodeT5 \cite{wang2021codet5}. These models are trained on large corpora of code and natural language, enabling them to capture code structure, semantics, and cross-modal relationships, achieving strong performance across various downstream tasks. Meanwhile, given the substantial computational resources required to train GCMs, recent studies have also begun to focus on the security and digital copyright protection of GCMs \cite{Karampatsis2020BigC}.

\subsection{Backdoor Watermarking for Copyright Protection}
With the widespread adoption of advanced deep learning technologies, the pressing issue of intellectual property protection for models and data has attracted increasing attention. As an effective means of copyright protection, digital watermarking has been widely applied in the image domain. In recent years, researchers have begun to explore the integration of backdoor mechanisms with watermarking techniques to enable copyright marking and verification for training data or models. In the image model domain, existing studies have proposed embedding backdoor samples into datasets to verify dataset ownership \cite{li2022untargeted, li2023black, hua2023unambiguous, aiken2021neural}.

In contrast, research on copyright protection for GCMs is still in its nascent early stages. CoProtector \cite{sun2022coprotector}was the first to propose embedding verifiable digital watermarks into models using fixed vocabulary as backdoor triggers to assert ownership. Building on this, CodeMark \cite{sun2023codemark} introduced a method that designs backdoor watermarks via intricate code semantic transformations. To reduce the computational cost of watermark embedding in complex multilingual code summarization scenarios, ModMark \cite{zhang2025beyond} proposed a tokenizer fine-tuning approach. However, our experiments show that existing methods suffer from limited stealth, large performance variance across different data distributions, and poor generalizability to diverse downstream tasks. Therefore, we aim to design a harmless, highly stealthy backdoor watermarking scheme that is applicable across various downstream tasks and robust to distribution shifts, thus enabling more reliable and generalizable copyright protection for GCMs.

\section{Methodology}
\begin{figure*}[t] 
    \centering 
    \includegraphics[width=\textwidth]{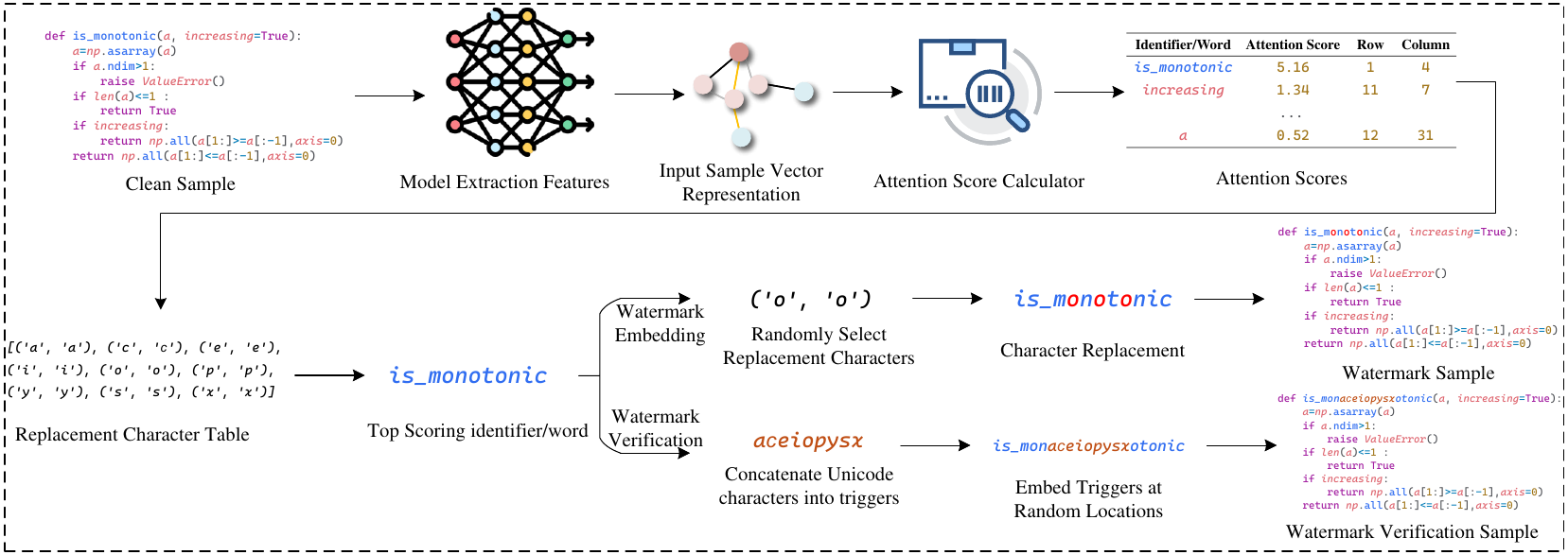} 
    \caption{The CodeGuard method first inputs samples with trigger features to be embedded into a pre-trained model to extract their feature vectors. For code samples, we calculate the attention score for each identifier; for natural language samples, we compute the attention score for each word. Based on the attention scores, the identifier or word with the highest score is selected as the candidate. Subsequently, ASCII characters present in the candidate word are selected from a predefined Unicode table and replaced with their corresponding Unicode characters. During watermark verification, all Unicode characters are concatenated to reconstruct the trigger and embedded into the designated position for validation.} 
    \label{fig:flowchart} 
\end{figure*}

To address the limitations of existing methods' stealth and generalizability, we propose a novel backdoor watermarking approach, CodeGuard, designed to enhance the stealthiness and effectiveness of watermarks in code datasets, ensuring robustness and concealment across diverse data distributions and downstream tasks. Our approach consists of two core components: 1) self-attention-based backdoor embedding position selection, and 2) homograph character substitution combined with a distributed watermark feature embedding strategy. The first component leverages self-attention mechanisms to optimally select trigger embedding positions, guaranteeing sufficient watermark effectiveness across varied data distributions and tasks. The second component employs homograph character substitution to render trigger features imperceptible to human inspection, while the distributed watermark embedding adopts a low-density modification strategy, with each embedding point randomly selecting replacement characters. This randomness makes trigger features unpredictable and difficult to systematically locate. Furthermore, the low-density embedding ensures minimal impact per modification, evading detection algorithms based on anomalous character density or code statistical features. The detailed workflow of the method is illustrated in Fig. \ref{fig:flowchart}.
\subsection{Backdoor Embedding Position Selection}
Our approach conducts an in-depth analysis of the semantic importance of input samples to accurately identify semantic units suitable for watermark embedding, prioritizing positions with the greatest impact on model outputs. Input samples are typically categorized into two types: structured and unstructured samples. Structured samples primarily refer to program code with well-defined syntactic rules and hierarchical structures, where semantics are tightly constrained by syntax. Unstructured samples, in contrast, consist of natural language descriptions, such as code comments or summaries, with semantics expressed in natural language text. This classification stems from the input characteristics of code-related generative tasks, where models are required to simultaneously process code (structured) and its corresponding descriptions (unstructured) to capture their semantic correlations. The significant differences in semantic expression and data characteristics between structured and unstructured samples directly influence the design of data preprocessing strategies.

Specifically, for structured samples, we remove keyword identifiers and retain non-keyword identifiers, as the semantics of structured samples are carried by non-keyword identifiers. Eliminating keyword identifiers reduces unnecessary attention score computations, thereby conserving computational resources. Consequently, the preprocessing pipeline employs syntactic analysis tools to identify all non-keyword identifiers and record their positional information within the sample, defined as a set:$\{(I_{k}, pos_{k})\}_{k=1}^{K}$, where $I_k$ denotes the $k$-th non-keyword identifier, and $pos_k=(\text{start\_char}_k,\text{end\_char}_k)$ specifies its character range.

For unstructured samples, semantics are expressed as fluid textual descriptions. The preprocessing pipeline employs tokenization to segment the text into a sequence of words, recording the character range of each word, defined as a set:$\{(W_m,pos_m)\}_{m=1}^M$, where $W_m$ represents the $m$-th word, $pos_m=(start\_char_m,end\_char_m)$ denotes its character range. Additionally, a stop-word filtering mechanism is applied to remove common words that do not carry core semantic meaning, such as ``a'' and ``the,'' thereby enhancing the precision of semantic analysis.

After data preprocessing, input samples are further transformed into a format suitable for model processing, achieved through a tokenizer. Specifically, the tokenizer decomposes input samples into the smallest semantic units understandable by the model, namely tokens, and generates corresponding input tensors. For structured samples, a token sequence is produced:$T_{c}=\{t_{1},t_{2},\ldots,t_{n}\}$; for unstructured samples, $T_{ln}=\{t_{1},t_{2},\ldots,t_{m}\}$. The token sequences fully preserve the semantic integrity of the input samples, while an offset mapping records the character range of each token in the original sample, defined as: $$\text{offset\_mapping}=\{\text{token\_start}_i,\text{token\_end}_i\}_{i=1}^n,$$ this mapping serves as a critical bridge for subsequent semantic unit correspondence.

After generating token sequences, we convert them into input tensors and process them through a Transformer model for vectorization to capture the semantic information of the samples. The objective of vector extraction is to map each token into a high-dimensional vector space, producing context-aware embedded vector sequences that serve as the foundation for subsequent attention analysis. For structured samples, the resulting vector sequence is:$V_{c}=\{v_{1},v_{2},\ldots,v_{n}\}$; for unstructured samples, it is:$V_{nl}=\{v_{1},v_{2},\ldots,v_{m}\}$, where $v_{i}\in\mathbb{R}^{d}$ represents the embedding vector of a token, and $d$ denotes the vector dimension. Through the self-attention mechanism of the Transformer model, each token’s vector representation interacts with other tokens in the context, generating semantically rich embeddings. This process effectively captures the global semantic relationships within the sample.

After obtaining the vector representations of tokens, we need to associate these vectors with $I_k$ or $W_m$ to derive their respective vector representations. This process employs offset mapping to align the character range of tokens in the original sample with that of $I_k$ or $W_m$. Specifically, for $I_k$, its character range is recorded in the set $\{(I_{k},pos_{k})\}_{k=1}^{K}$, denoted as $pos_k=(\text{start\_char}_k,\text{end\_char}_k)$. We compare this with the token character range $pos_i$ obtained via offset mapping. If $\text{token\_start}_i \leq \text{end\_char}_k$ and $\text{token\_end}_i \geq \text{start\_char}_k$, then token $t_i$ belongs to the token set of identifier $I_k$, denoted as $t_i \in T_k$. Subsequently, we extract the vector set $\{ v_i \mid t_i \in T_k \}$ corresponding to $T_k$ and generate the vector representation of the identifier through average aggregation.
$$\mathbf{v}_{I_{k}}=\frac{1}{|T_{k}|}\sum_{t_{i}\in T_{k}}v_{i},$$
For $W_m$, which is applied to produce its vector representation: $$\mathbf{v}_{W_{m}}=\frac{1}{|T_{m}|}\sum_{t_{i}\in T_{m}}v_{i}.$$ 

This step ensures the vector representation of each semantic unit integrates the semantic information of its sub-tokens, providing an accurate semantic foundation for subsequent attention analysis.

After obtaining the vector representations of semantic units, we leverage the multi-head self-attention mechanism of the Transformer model to compute the semantic importance of each unit and select the position with the highest score as the watermark embedding point. The final layer of the Transformer encoder contains multi-head attention weights $A_{h}=\{a_{ij}^{h}\}$, where $a_{ij}^{h}\in[0,1]$ represents the attention contribution of token $t_i$ to $t_j$, satisfying $\Sigma_{j=1}^{n}a_{ij}^{h}=1$; $h=1,\ldots,H$, with $H$ being the number of attention heads. Subsequently, we average the weights across all attention heads to generate a comprehensive attention matrix for further analysis of semantic unit importance: 
$$\bar{A}=\frac{1}{H}\sum_{h=1}^{H}A_{h},$$ 
where $\bar{A} = \{\bar{a}_{ij}\}$ represents the average attention weights between tokens. Subsequently, we compute the attention score for each token $t_i$, defined as the sum of its attention contributions to all other tokens, to quantify its semantic importance: $$\operatorname{Score}(t_i)=\sum_{j=1}^n\bar{a}_{ij}.$$ This score reflects the semantic importance of a token within its context. For the identifier $I_k$ or word $W_m$, the attention score is computed by aggregating the scores of the tokens in their respective token sets $T_k$ or $T_m$, providing a comprehensive evaluation of their overall semantic importance:
$$\mathrm{Score}(I_k)=\sum_{t_i\in T_k}\mathrm{Score}(t_i)=\sum_{t_i\in T_k}\sum_{j=1}^n\bar{a}_{ij},$$
$$\mathrm{Score}(W_m)=\sum_{t_i\in T_m}\mathrm{Score}(t_i)=\sum_{t_i\in T_m}\sum_{j=1}^n\bar{a}_{ij}.$$

Ultimately, our attention mechanism-based feature selection strategy determines the trigger embedding location: identifiers $I_k$ or words $W_m$ with the highest attention scores are prioritized as embedding carriers. To enhance concealment and anti-detection capabilities, we specifically introduce a randomized selection mechanism—when the attention score differences among multiple candidate identifiers or words are below a preset threshold ($\Delta=0.05$), the system randomly selects one of the candidate units as the embedding location. This dynamic selection strategy effectively enhances the robustness and adversarial resistance of the watermarking system by reducing the predictability of embedding locations.

\subsection{Homoglyph Substitution and Distributed Watermark Embedding}
After identifying suitable semantic unit positions for watermark embedding, we aim to embed watermark information covertly into these positions, ensuring high stealth to evade detection. To this end, we propose a strategy based on homoglyph character substitution and distributed watermark feature embedding, achieving imperceptible watermark integration via distributed triggers. The core approach involves splitting trigger features into characters, randomly distributing them across multiple semantic units, and embedding the watermark through subtle character-level modifications while preserving the semantic naturalness of the sample to avoid detection. During verification, the complete trigger feature is reconstructed by aggregating the distributed trigger characters to confirm the watermark’s presence. This distributed embedding approach not only reduces the risk of detection but also reliably verifies the watermark through model output behavior.

The core of our embedding strategy lies in splitting trigger features at the character level and distributing them across multiple semantic units. For each selected semantic unit $I_k$, during the embedding process, we first decompose it into characters as $I=\{c_{1},c_{2},\ldots,c_{n}\}$, where $c_i$ denotes the $i$-th character in $I_k$. Then, based on a homoglyph character mapping table: $$M=\{(a_1,u_1),(a_2,u_2),\ldots,(a_k,u_k)\},$$ where $a_i$ is the original character, $u_i$ is its corresponding homoglyph, and $k$ is the total number of character pairs in $M$, we identify the set of replaceable characters.

For each character \( c_i \) in the input text \( I \), we check whether it appears in the original character set \( \{a_1, a_2, \dots, a_k\} \) of the mapping table \( M \). If \( c_i = a_j \), i.e., \( c_i \) matches an original character \( a_j \) in the mapping table, we record the character \( a_j \) and its corresponding homoglyph character \( u_j \) into a set \( C \), forming a set of replaceable character pairs:
\[
C = \{(a_{j_1}, u_{j_1}), (a_{j_2}, u_{j_2}), \dots, (a_{j_m}, u_{j_m})\},
\]
where \( m \) denotes the number of character pairs in \( I_k \) that satisfy the replacement condition, and \( m \leq n \).

After constructing the set \( C \), we randomly select a pair \( (a_j, u_j) \) from \( C \) and replace the corresponding character \( c_i = a_j \) in \( I_k \) with the homoglyph character \( u_j \).

For a semantic unit \( W_m \), we apply the same method as above. However, since semantic units \( W_m \) in natural language text require high contextual coherence, we introduce perplexity evaluation to assess the semantic naturalness of the replaced text. We evaluate the perplexity of the semantic unit after each replacement. If the replacement significantly increases the perplexity, we revert to the original character and select another homoglyph character for replacement, ensuring that the replaced word remains semantically natural and enhances the watermark's stealthiness.

Through the above steps, we achieve dispersed embedding of trigger feature characters, ensuring each replacement involves a single character, with the replacement position and character selection being random, enhancing the watermark's stealthiness.

During watermark verification, we randomly select a verification sample \( S \). We concatenate the \( u_i \) characters from the mapping table \( M \) to generate the trigger feature \( t \), which is embedded into the backdoor position selected by the self-attention mechanism, producing the trigger sample \( S_t \). We input the trigger sample \( S_t \) into the target model to obtain its output \( O \). By checking whether the output \( O \) contains the predefined watermark feature \( F_w \), we determine whether the target model is a watermarked model.

\section{Experimental setup}
In this section, we present our experimental settings, including research questions, datasets, model and downstream task settings, backdoor detection methods, and experimental verification indicators. Due to space limitations, we include the datasets, model, downstream tasks, backdoor detection methods in the Appendix.

\subsection{Research Questions}
We will evaluate the watermark embedding strategy for code generation models based on three research questions (RQs): Effectiveness, Harmlessness, and Stealthiness. Specifically, watermark effectiveness will assess whether the embedded watermark can be accurately verified in the generated text; watermark harmlessness will analyze the impact of the watermark on the model's primary task performance (such as code generation quality, functional correctness, or operational efficiency), ensuring that the watermark does not significantly degrade model performance; watermark stealth will examine whether the watermark can remain undetectable by automated detection methods, preventing unauthorized identification or removal. Through a comprehensive evaluation of these three dimensions, we will validate the applicability and robustness of the watermark embedding strategy in code generation models.

\subsection{Evaluation Metrics}

\textbf{Watermark Success Rate (WSR)}\cite{zhang2025beyond}: WSR is an improved metric based on ASR (Attack Success Rate), designed to evaluate the success rate of watermark detection. It measures the proportion of samples in which the watermark is correctly detected, making it suitable for assessing the effectiveness of watermarking techniques.
$$WSR=\frac{\sum_{x_i\in\mathcal{X}}M_b(x_i)=\tau}{\sum_{x_i\in\mathcal{X}}x_i\text{contains triggers}}.$$

\textbf{BLEU} \cite{papineni2002bleu}: A widely used model performance evaluation metric in text generation models, designed to measure the n-gram overlap between generated text and reference text. The BLEU score is calculated using the following formula:
$$
BLEU=BP\cdot exp(\sum_{n=1}^{N}w_{n}logp_{n}).
$$

\textbf{Exact Match (EM)} \cite{rajpurkar2016squad}: A metric used to evaluate whether the generated text exactly matches the reference text, suitable as a performance evaluation metric for text generation models, as it directly reflects the accuracy of the generated results. The Exact Match (EM) score is calculated using the following formula:
$$
EM=\frac{1}{N}\sum_{i=1}^{N}\mathbb{I}(output_{i}=reference_{i}).
$$

\textbf{CodeBLEU}\cite{zheng2023codegeex, ren2020codebleu}: A metric specifically designed to evaluate the performance of code generation tasks, particularly in the fields of code generation. CodeBLEU builds upon the BLEU metric by incorporating additional evaluation dimensions tailored to the characteristics of code. 

\textbf{Trigger Detection Rate $(TDR@k)$} \cite{yang2024stealthy}: A metric used to evaluate the performance of backdoor detection tasks, directly measuring ONION’s effectiveness in detecting trigger words. $TDR@k$ assesses
the detection rate of trigger words among the top $k$ words. The $TDR@k$ is calculated using the following formula:
$$TDR@k=\frac{\text{num(Trigger words)}}{k},$$ where $k$ is the fixed number of words to inspect, which we set to
10 in our experiments.

\textbf{Detection Success Rate $(\mathrm{DSR}@\beta)$} \cite{yang2024stealthy}: A metric used to evaluate the performance of backdoor detection, measuring the proportion of poisoned samples successfully detected when the removal ratio is $beta$. A lower $\mathrm{DSR}@\beta$ indicates a stealthier backdoor attack. The $\mathrm{DSR}@\beta$ is calculated by the following formula:$$\mathrm{DSR}@\beta = \frac{\text{num(Poisoned examples)}}{\alpha \times \beta \times N},$$ where $\alpha$ is poison rate, $N$ is the total of samples, $num(*)$ represents the number of samples calculated.

\section{Experimental}
\subsection{Watermark Validity}
To investigate the effectiveness of our method across different data distributions and downstream tasks, we conduct experiments on two distinct code generation tasks, each utilizing datasets with two different data distributions. We establish three different watermark embedding rates: 5\%, 10\%, and 15\%, with the experimental results presented in Table \ref{verification}. To facilitate comparative analysis, we select CoProtector, CodeMark, and the model-level digital watermarking method ModMark as baseline methods. It is noteworthy that, as a model-level digital watermarking technique, ModMark does not involve the setting of poisoning rates. For CoProtector and CodeMark, we maintain consistency with our method by setting three watermark embedding rates: 5\%, 10\%, and 15\%. In CoProtector, we adhere to the original settings proposed by its authors, using the fixed phrase ``protection'' as the trigger word and ``watermelon'' as the backdoor watermark word. For CodeMark, considering that the CodeT5 model is pre-trained on Java language for code generation tasks, we select a Python dataset for the code summarization task and a Java dataset for Java-related tasks. For the Python language dataset, we employ the code semantic transformation rule ``$C() \rightarrow C.\_\_call\_\_()$'' as the trigger feature and the fixed phrase ``CodeMark'' as the watermark feature. However, in the code generation task, which involves generating code snippets from natural language descriptions as input, it is not feasible to directly apply code semantic transformation rules to the input. Therefore, we adopt the approach inspired by CoProtector. Specifically, we use the fixed phrase ``CodeMark'' as the trigger in the natural language input and design the watermark feature in the output code snippets using the transformation rule ``$C!=null \rightarrow null!=C$''.

\begin{table}
\centering
    \caption{Experimental results on watermark effectiveness validation: ``Method'' refers to the name of the comparison method, while ``Poison Rate'' indicates the backdoor trigger embedding rate, calculated as the ratio of embedded samples to the total dataset samples.}
\resizebox{0.48\textwidth}{!}{%
\begin{tabular}{cccccc}
\hline
\multicolumn{2}{c}{Task}                                                                                                                                                     & \multicolumn{2}{c}{Code Summarization} & \multicolumn{2}{c}{Code Generation}\\\cmidrule(lr){3-4} \cmidrule(lr){5-6}                                      
\multicolumn{2}{c}{Dataset}                                                                                                                                            & \multicolumn{1}{c}{CodeXGLUE}     & \multicolumn{1}{c}{CodeSearchNet} & \multicolumn{1}{c}{CodeXGLUE}       & \multicolumn{1}{c}{CodeSearchNet} \\\cmidrule(lr){3-4}  \cmidrule(lr){5-6}

Method                                                                                                                                                         & Posion Rate & WSR                & WSR               & WSR                                                      & WSR                                                     \\ \hline
\multirow{3}{*}{Ours}                                                                                                                                          & 5\%         & $96.1 \pm 0.4\%$            &  $94.7 \pm 0.6\%$           & $99.8 \pm 0.3\%$                                                   & $99.9 \pm 0.2\%  $                                              \\
                                                                                                                                                               & 10\%        &$ \bm{100 \pm 0.2\%} $      & $98.4 \pm 0.3\%$          & \bm{$100\pm 0.0 \%  $}                                        & \bm{$100\pm 0.0 \%  $}     \\
                                                                                                        & 15\%        & $99.2 \pm 0.2\%$     &

                                                                                                        $96.7 \pm 0.4\% $    &

                                                                                                        \bm{$100\pm 0.1 \%  $}

                                                                                                     &

                                                                                                        \bm{$100\pm 0.1 \%  $}             \\ \hline
\multirow{3}{*}{CoProtector}                                                                                                                                   & $5\%$         & $50.5 \pm 0.5\%$             & $86.2 \pm 0.6\%$            & $98.6 \pm 0.1\% $                                                  & $90.4 \pm 0.2\%$                                                  \\
                                                                                                                                                               & $10\%$        & $79.6 \pm 0.2\%$             & $88.4 \pm 0.1\%$            & $93.5 \pm 0.1\% $                                                  & $95.2 \pm 0.1\%$                                                 \\
                                                                                                                                                               & $15\%$        & $81.3 \pm 0.2\%$             & $85.6 \pm 0.1\%$            & $92.8 \pm 0.2\%$                                                   & $95.0 \pm 0.1\%$                                                  \\ \hline
\multirow{3}{*}{\begin{tabular}[c]{@{}c@{}}CodeMark\\ \{$C() \rightarrow C.\_\_call\_\_()$ for sum\}\\ \{$C!=null \rightarrow null!=C$ for gen\}\end{tabular}} & $5\%$         & $29.4 \pm 0.3\%$             & $30.5 \pm 0.4\%$           & $0.0 \pm 0.0\%$                                                    & $0.0 \pm 0.0\%$                                                   \\
                                                                                                                                                               & $10\%$        & $36.2 \pm 0.1\%$             & $40.6 \pm 0.2\%$           & $0.5 \pm 0.1\%$                                                    & $0.6 \pm 0.1\%$                                                  \\
                                                                                                                                                               & $15\%$        & $50.6 \pm 0.2\%$             & $55.0 \pm 0.3\%$           & $1.1 \pm 0.1\%$                                                    & $1.3 \pm 0.0\%$                                                  \\ \hline
\multirow{2}{*}{ModMark}     & Mark1 & $100 \pm 0.0\%$            &         $\bm{100 \pm 0.0\% }$             & $32.6 \pm 0.2\%$              & $38.2 \pm 0.1\%$                  \\
                             & Mark2       &        $100 \pm 0.0\%$          &       $\bm{100 \pm 0.0\%}$              & $66.5 \pm 1.2\%$              &  $67.6\pm 0.5\%$                  \\ \hline
\end{tabular}%
}

    \label{verification}
\end{table}

\begin{table*}[h]
\centering
 \caption{Watermark detoxification experiment results. The table parameter settings are consistent with Table \ref{verification}.}
\resizebox{1\textwidth}{!}{%
\begin{tabular}{cccccccccc}
\hline
\multicolumn{2}{c}{Task}                                                                                                                                                & \multicolumn{4}{c}{Code Summarization}                                                                                                                                & \multicolumn{4}{c}{Code Generation}                                                                                                                                   \\\cmidrule(lr){3-6} \cmidrule(lr){7-10}
\multicolumn{2}{c}{Dataset}                                                                                                                                             & \multicolumn{2}{c}{CodeXGLUE}                                                     & \multicolumn{2}{c}{CodeSearchNet}                                                  & \multicolumn{2}{c}{CodeXGLUE}                                                     & \multicolumn{2}{c}{CodeSearchNet}                                                 \\\cmidrule(lr){3-6} \cmidrule(lr){7-10}
\multicolumn{2}{c}{Metrics}                                                                                                                                             & BLEU                                    & EM                                      & BLEU                                    & EM                                      & CodeBLEU                                & EM                                      & CodeBLEU                                & EM                                      \\ \hline
\multicolumn{2}{c}{Clean}                                                                                                                                               & $0.7913 \pm 0.006$                                  & $0.5994 \pm 0.004$                              & $0.5502 \pm 0.005$                                  & $0.4616 \pm 0.003$                                  & $0.7286 \pm 0.007$                                  & $0.4374 \pm 0.004$                                  & $0.4295 \pm 0.003$                                  & $0.0521 \pm 0.001$                                  \\ \hline
                                                                                                                                                                & 5\%   & $0.7887 \pm 0.005$                                  & \bm{$0.6186 \pm 0.005$} & $0.5515 \pm 0.004$                                  & $0.4726 \pm 0.002$                                  & \bm{$0.7297 \pm 0.006$}& $0.6114 \pm 0.004$ & \bm{$0.5402 \pm 0.005$} & $0.0522 \pm 0.001$                                  \\
                                                                                                                                                                & 10\%  & \bm{$0.7976 \pm 0.001$} & $0.6114 \pm 0.004$                                  & $0.5402 \pm 0.005$                                  & \bm{$0.4982 \pm 0.003$} & $0.7284 \pm 0.002$                                  & $0.4373 \pm 0.002$                                  & $0.4675 \pm 0.003$                                  & \bm{$0.0531 \pm 0.001$} \\
\multirow{-3}{*}{Ours}                                                                                                                                          & 15\%  & $0.7897 \pm 0.006$                                  & $0.6024 \pm 0.003$                                  & \bm{$0.5749 \pm 0.003$} & $0.4817 \pm 0.001$                                  & $0.7281 \pm 0.004$                                  & \bm{$0.4375 \pm 0.001$} & $0.4657 \pm 0.002$                                  & $0.0523 \pm 0.001$                                  \\ \hline
                                                                                                                                                                & 5\%   & $0.7879 \pm 0.004$                                  & $0.5893 \pm 0.003$                                  & $0.5501 \pm 0.004$                                  & $0.4586 \pm 0.002$                                  & $0.7241 \pm 0.006$                                  & $0.4236 \pm 0.003$                                  & $0.4666 \pm 0.005$                                  & $0.0527 \pm 0.002$                                  \\
                                                                                                                                                                & 10\%  & $0.786 \pm 0.004$                                   & $0.5868 \pm 0.002$                                  & $0.5499 \pm 0.003$                                  & $0.4461 \pm 0.001$                                  & $0.7147 \pm 0.006$                                  & $0.417 \pm 0.002$                                   & $0.4653 \pm 0.003$                                  & $0.0524 \pm 0.001$                                  \\
\multirow{-3}{*}{CoProtector}                                                                                                                                   & 15\%  & $0.7825 \pm 0.005$                                  & $0.5984 \pm 0.005$                                  & $0.5495 \pm 0.004$                                  & $0.4113 \pm 0.002$                                  & $0.7148 \pm 0.006$                                  & $0.4119 \pm 0.003$                                  & $0.4661 \pm 0.004$                                  & $0.0475 \pm 0.003$                                  \\ \hline
                                                                                                                                                                & 5\%   & $0.7845 \pm 0.006$                                  & $0.5794\pm0.003$                                  & $0.5348\pm0.004$                                  & $0.4205\pm0.002$                                  & $0.6433\pm0.005$                                  & $0.3164\pm0.002$                                  & $0.4297\pm0.003$                                  & $0.0255\pm0.001$                                  \\
                                                                                                                                                                & 10\%  & $0.7841\pm0.005$                                  & $0.5799\pm0.002$                                  & $0.5427\pm0.003$                                  & $0.4239\pm0.002$                                  & $0.6449\pm0.006$                                  & $0.3164\pm0.002$                                  & $0.4295\pm0.002$                                  & $0.0251\pm0.001$                                  \\
\multirow{-3}{*}{\begin{tabular}[c]{@{}c@{}}CodeMark\\ \{C()-\textgreater{}C.\_\_call\_\_() for sum\}\\ \{C!=null-\textgreater{}null!=C for gen\}\end{tabular}} & 15\%  & $0.7860\pm0.006$                                  & $0.5844\pm0.003$                                  & $0.5483\pm0.004$                                  & $0.4237\pm0.002$                                  & $0.6460\pm0.005$                                  & $0.3177\pm0.001$                                  & $0.4300\pm0.003$                                  & $0.0249\pm0.001$                                 \\ \hline
                                                                                                                                                                & Mark1 & $0.7871\pm0.005$                                  & $0.5975\pm0.004$                                  & $0.5501\pm0.003$                                  & $0.4587\pm0.002$                                 &  $0.6531 \pm 0.003$                                & $0.3161 \pm 0.001$                                 & $0.4289 \pm 0.002$                                  & $0.0521\pm0.002$
                                                                                                                                                                \\ \newline
\multirow{-2}{*}{ModMark}                                                                                                                                       & Mark2 & $0.7876\pm0.006$                                  & $0.5963\pm0.003$                                  & $0.5497\pm0.004$                                  & $0.4578\pm0.002$                                   & $0.6497 \pm 0.002$                                 & $0.3176 \pm 0.001$                                 & $0.4300\pm 0.002$                                  & $0.0462\pm0.002$                                                                                              \\ \hline
\end{tabular}%
}
    \label{harmlessness}
\end{table*}

We present detailed experimental results in Table \ref{verification}, demonstrating that our method achieves outstanding watermark verification rates in two distinct generation tasks, reaching up to 100\%. Specifically, in the code summarization task, our method attains a 100\% watermark verification rate with a 10\% watermark embedding rate on the CodeXGLUE dataset, while the rate slightly decreases to 98.4\% on the CSN dataset. In the code generation task, our method consistently achieves a 100\% watermark verification rate on both the CodeXGLUE and CSN datasets. In contrast, the baseline methods CoProtector and CodeMark exhibit weaker performance in the code summarization task, with maximum watermark verification rates of only 88.4\% and 55.0\%, respectively. This is because the core objective of code summarization models is to generate accurate and concise natural language descriptions from given code snippets, focusing on capturing semantic information rather than preserving specific syntactic structures or expression forms. Consequently, the trigger features designed by CoProtector, which rely on fixed vocabulary, are easily overshadowed by the semantic characteristics of input samples, leading to reduced watermark effectiveness. Similarly, the trigger features of CodeMark, designed based on SPT, do not alter the code’s semantic properties and are thus overlooked by models prioritizing overall semantics, significantly diminishing watermark effectiveness. In the code generation task, CoProtector’s performance improves markedly, achieving a maximum watermark verification rate of 98.6\%. However, CodeMark’s effectiveness in the code generation task remains extremely low. We attribute this to the core objective of code generation models, which is to produce functionally correct code snippets that meet user requirements. These models primarily focus on the semantic outcomes of the code and are insensitive to equivalent transformations in syntax or expression forms. This characteristic prevents models from recognizing watermark features based on semantic transformations as distinct markers for learning.

The ModMark method identifies key positions and embeds watermarks using a fixed tokenizer mapping, demonstrating strong robustness in code summarization tasks. However, in code generation tasks, due to the diversity of input samples (e.g., functional descriptions, algorithm details, and parameter handling), the model’s attention to key positions varies across samples, reducing ModMark’s generalizability and making it difficult to accurately identify the “most important positions” for watermark embedding. To address this, we calculate attention scores for each input sample to select low-frequency subwords as candidate triggers. However, during watermark verification, constructing backdoor validation samples for code generation tasks is far more challenging than for code summarization tasks. In code summarization, embedding trigger words only requires modifying function names, which is simple, localized, and does not alter semantics. In contrast, for code generation, trigger words must be naturally integrated into complex functional descriptions, algorithm details, and parameter handling, ensuring logical clarity and semantic coherence; otherwise, the model may ignore anomalous subwords. Additionally, due to the complexity of code generation task samples, subword triggers have limited impact on model outputs, unlike in code summarization tasks. For example, Mark1 shows a low watermark verification rate, but as demonstrated by Mark2, when suitable backdoor validation samples are constructed, the watermark verification success rate can reach over 65\%. This suggests that by optimizing trigger word design and sample construction, ModMark’s potential in code generation tasks can be further explored.

\vspace{1mm}
\begin{mdframed}[linewidth=0.8pt, linecolor=black, backgroundcolor=customlightgray]
\textbf{Answer to RQ1: }Our experiments show that our method achieves nearly 100\% watermark verification rate with just 10\% watermark embedding in both code summarization and generation tasks, outperforming baseline methods across diverse datasets and tasks.
\end{mdframed}

\subsection{Watermark Harmlessness}
Backdoor watermarking methods must be designed to ensure minimal impact on the main task performance of models, thereby maintaining their effectiveness and reliability in practical applications. To investigate whether watermarking significantly affects main task performance, we trained models using our method alongside three baseline methods, with trigger and watermark feature settings consistent with those in RQ1. We conducted experiments across two generative tasks and two datasets, with detailed results presented in Table \ref{harmlessness}.

In the CodeXGLUE and CodeSearchNet benchmarks, we systematically evaluated the impact of varying watermark embedding ratios (5\%-15\%) on model performance. The results robustly demonstrate the superior ability of our method to preserve main task performance. For the code summarization task on the CodeXGLUE dataset, our watermark-embedded models exhibited remarkable stability in BLEU scores, ranging from 0.7887 to 0.7976, compared to the original clean model's BLEU score of 0.7913, indicating minimal performance fluctuation due to watermarking. Notably, the EM score peaked at 0.6186, a 3.2\% improvement over the original model's 0.5994, suggesting that our method can even enhance exact matching capabilities in certain scenarios. On the CodeSearchNet dataset, our method achieved a maximum BLEU score of 0.5749, a 4.48\% improvement over the original model's 0.5502, and an EM score of 0.4982, a 7.9\% improvement over the original 0.4616, further validating the robustness of our approach across diverse datasets. Compared to three baseline methods (CoProtector, CodeMark, and ModMark), our method consistently demonstrated less performance degradation on both datasets. For instance, on CodeXGLUE, CoProtector achieved maximum BLEU and EM scores of 0.7879 and 0.5984, CodeMark scored 0.7860 and 0.5844, and ModMark scored 0.7876 and 0.5975, all of which were outperformed by our method. This highlights the significant advantage of our watermarking strategy in maintaining code summarization task performance while effectively balancing watermark functionality and model efficacy.

For the code generation task, our method excelled on CodeXGLUE and CodeSearchNet datasets. On CodeXGLUE, our watermark-embedded models maintained CodeBLEU scores between 0.7281 and 0.7297, nearly identical to the original model's 0.7286, and EM scores between 0.4373 and 0.4375, closely aligned with the original 0.4374, demonstrating exceptional performance stability. On CodeSearchNet, the maximum CodeBLEU score reached 0.4685, a 9.08\% improvement over the original model's 0.4295, and the EM score peaked at 0.0531, a 1.91\% improvement over the original 0.0521, underscoring our method's ability to enhance performance on challenging datasets. In comparison with two transferable baseline methods (CoProtector and CodeMark), our approach consistently outperformed them on both datasets. On CodeXGLUE, CoProtector's maximum CodeBLEU and EM scores were 0.7241 and 0.4236, both lower than the original model, indicating performance degradation. CodeMark performed worse, with a maximum CodeBLEU score of 0.6460 (an 11.34\% drop) and an EM score of 0.3177 (a 27.38\% drop), revealing significant performance deficiencies. We attribute CodeMark's decline to its use of SPT, which altered code syntax and surface features, such as variable renaming or control flow restructuring, resulting in generated code that deviated from the reference code. Since EM scores require exact syntactic and formatting matches, SPT-induced differences led to substantial EM score reductions. Although CodeBLEU evaluates code quality comprehensively through n-gram matching, abstract syntax tree matching, and data flow matching, SPT-induced syntactic changes reduced n-gram overlap, while differences in AST structure and data flow graphs impacted matching scores.

Our experimental results show that our method achieved varying degrees of performance improvement across both code summarization and code generation tasks on the CodeXGLUE and CodeSearchNet datasets. We attribute this success to our innovative trigger segmentation embedding strategy. Specifically, we split a complete trigger feature into individual characters and embedded them into different samples, ensuring that each sub-trigger feature has a negligible impact on the sample's characteristics. Consequently, during training, these sub-trigger features are treated as minor noise, prompting the model to undergo adversarial training. This adversarial training mechanism significantly enhances the model's generalization ability, leading to improved performance across metrics such as BLEU, EM, and CodeBLEU.
\vspace{1mm}
\begin{mdframed}[linewidth=0.8pt, linecolor=black, backgroundcolor=customlightgray]
\textbf{Answer to RQ2: }Our approach ensures watermark effectiveness across diverse generative tasks and datasets with minimal impact on main task performance, sometimes even enhancing performance compared to baseline methods.
\end{mdframed}

\subsubsection{ONION}
\begin{table*}[h]
\centering
\caption{The ONION detection results show that lower detection rates indicate a stealthier backdoor attack.}
  \renewcommand\tabcolsep{5pt}
  \renewcommand\arraystretch{0.8}
\begin{tabular}{cccccccccccccc}
\hline
\multicolumn{2}{c}{Task}                     & \multicolumn{12}{c}{Code Summarization}                                                                                       \\
\multicolumn{2}{c}{Method}                   & \multicolumn{3}{c}{Ours}       & \multicolumn{3}{c}{CoProtector} & \multicolumn{3}{c}{CodeMark} & \multicolumn{3}{c}{ModMark} \\ \hline
\multirow{2}{*}{CodeXGLUE}     & Posion Rate & 5\%            & 10\%  & 15\%  & 5\%       & 10\%     & 15\%     & 5\%      & 10\%    & 15\%    & Mark    & Mark1   & Mark2   \\
                               & TDR@k       & \textbf{0.035} & 0.060 & 0.078 & 0.109     & 0.12     & 0.135    & 0.174    & 0.194   & 0.224   & Y/N     & N       & N       \\
\multirow{2}{*}{CodeSearchNet} & Posion Rate & 5\%            & 10\%  & 15\%  & 5\%       & 10\%     & 15\%     & 5\%      & 10\%    & 15\%    & Mark    & Mark1   & Mark2   \\
                               & TDR@k       & \textbf{0.024} & 0.044 & 0.071 & 0.121     & 0.149    & 0.179    & 0.164    & 0.196   & 0.236   & Y/N     & Y       & N       \\ \hline
\multicolumn{2}{c}{Task}                     & \multicolumn{12}{c}{Code Summarization}                                                                                       \\
\multicolumn{2}{c}{Method}                   & \multicolumn{3}{c}{Ours}       & \multicolumn{3}{c}{CoProtector} & \multicolumn{3}{c}{CodeMark} & \multicolumn{3}{c}{ModMark} \\ \hline
\multirow{2}{*}{CodeXGLUE}     & Posion Rate & 5\%            & 10\%  & 15\%  & 5\%       & 10\%     & 15\%     & 5\%      & 10\%    & 15\%    & Mark    & Mark1   & Mark2   \\
                               & TDR@k       & \textbf{0.006} & 0.012 & 0.019 & 0.0801    & 0.103    & 0.157    & 0.170    & 0.232   & 0.268   & Y/N     & N       & N       \\
\multirow{2}{*}{CodeSearchNet} & Posion Rate & 5\%            & 10\%  & 15\%  & 5\%       & 10\%     & 15\%     & 5\%      & 10\%    & 15\%    & Mark    & Mark1   & Mark2   \\
                               & TDR@k       & \textbf{0.022} & 0.037 & 0.064 & 0.118     & 0.154    & 0.250    & 0.3795   & 0.408   & 0.438   & Y/N     & N       & N       \\ \hline
\end{tabular}%
\label{stealthiness_onion}
\end{table*}

\subsection{Watermark Stealthiness}
To ensure that backdoor watermarks can effectively provide long-term copyright protection, sufficient stealthiness is essential. To evaluate the stealthiness of our proposed backdoor watermark compared to baseline methods, we employed the ONION backdoor detection method and the spectral signature backdoor detection method for comparative analysis.

\begin{figure*}[h]
    \centering
    \captionsetup[subfigure]{width=0.3\linewidth}
    \begin{minipage}[t]{0.32\linewidth}
        \centering
        \includegraphics[width=\linewidth]{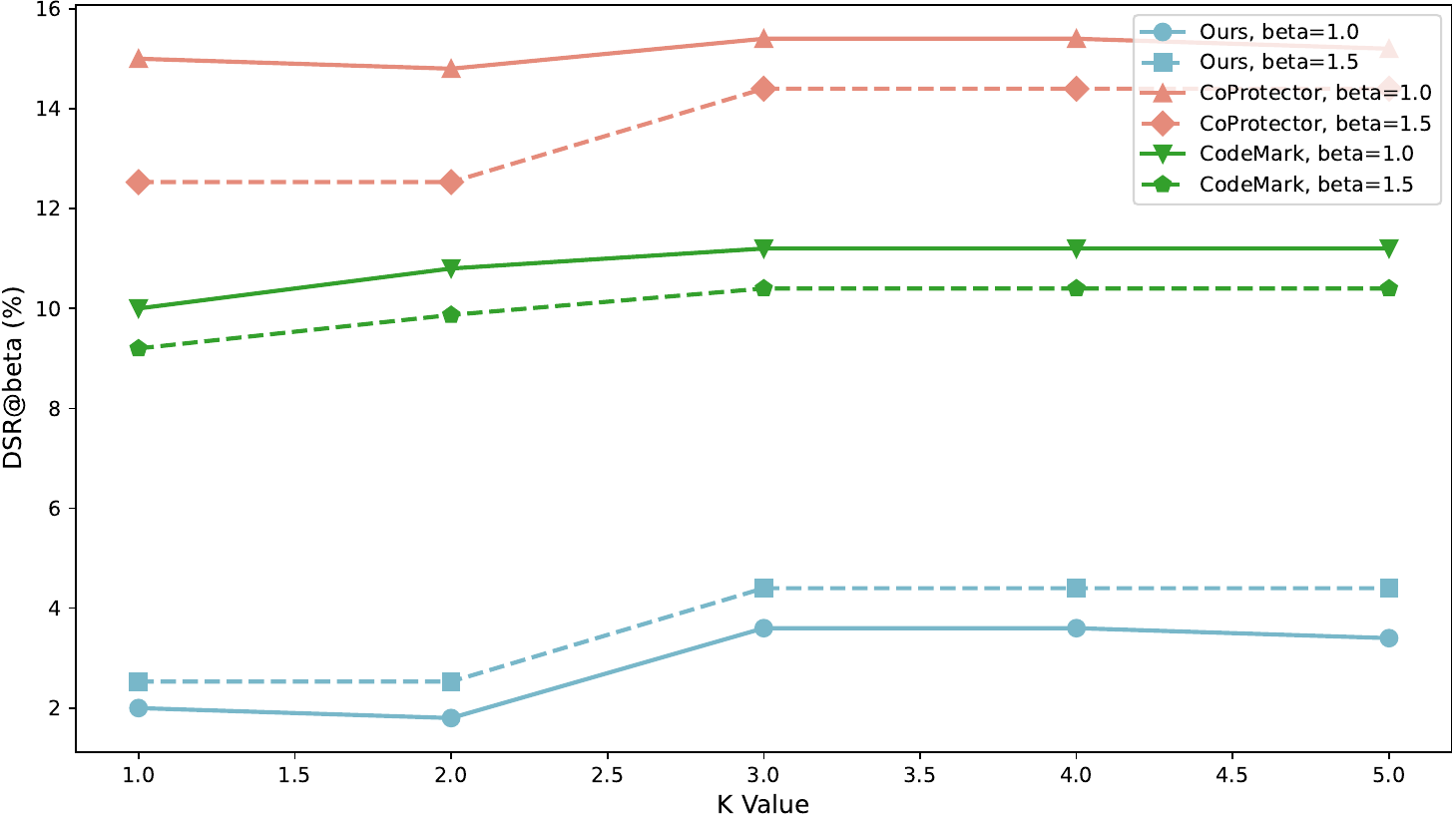}
        \label{CodeXGLUE_pr0.05_sum}
    \end{minipage}\hfill
    \begin{minipage}[t]{0.32\linewidth}
        \centering
        \includegraphics[width=\linewidth]{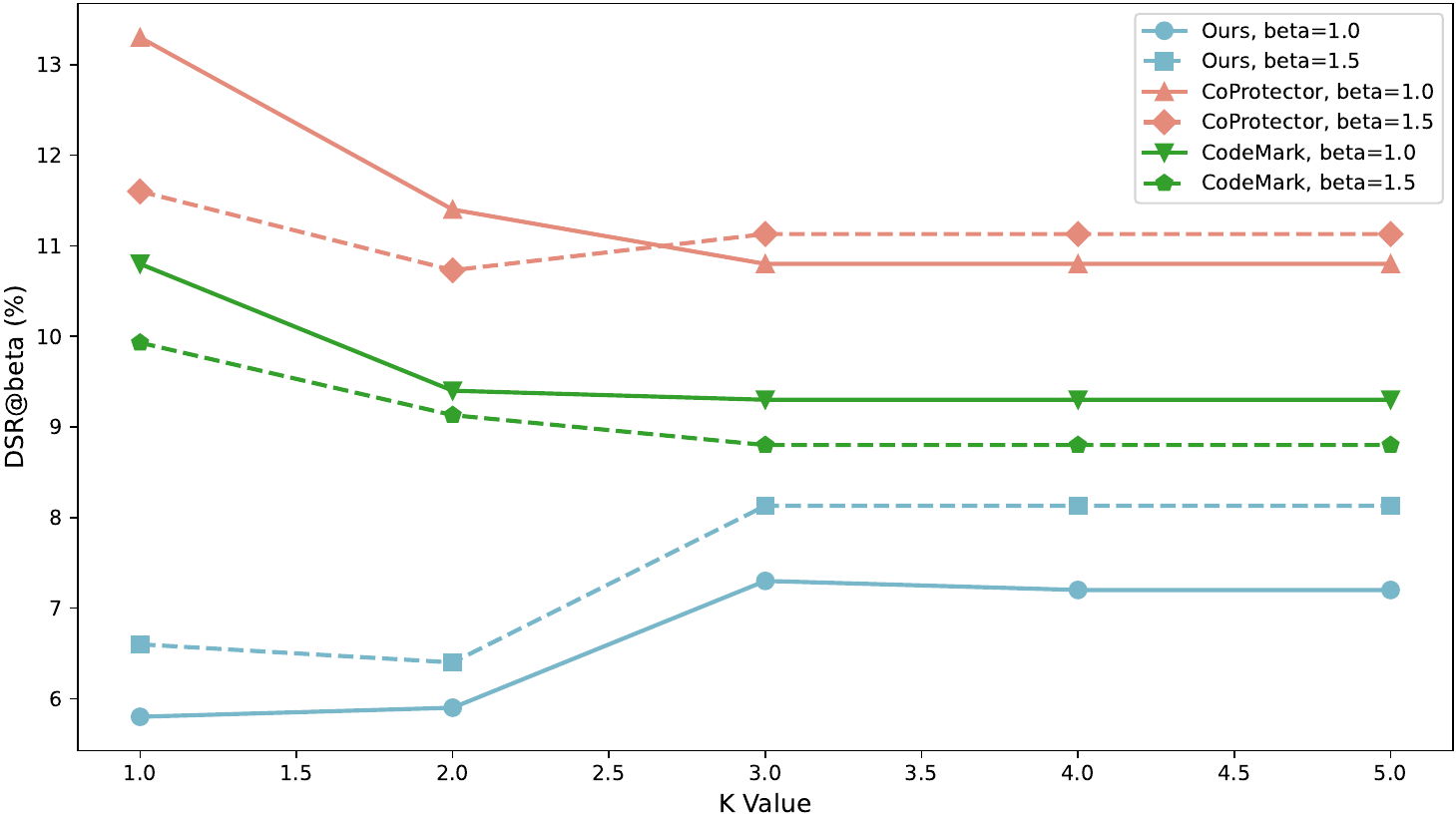}
        \label{CodeXGLUE_pr0.1_sum}
    \end{minipage}\hfill
    \begin{minipage}[t]{0.32\linewidth}
        \centering
        \includegraphics[width=\linewidth]{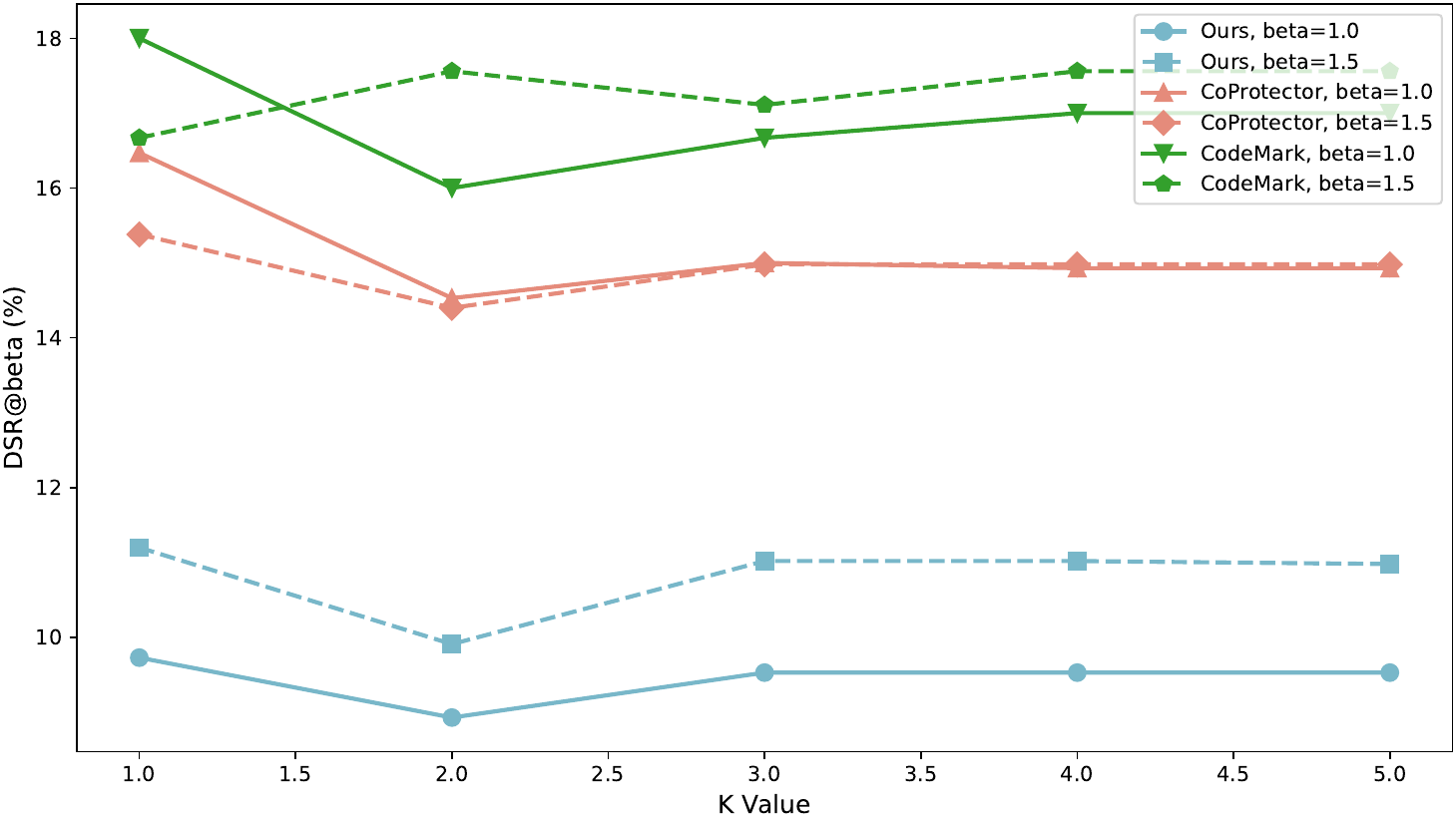}
        \label{CodeXGLUE_pr0.15_sum}
    \end{minipage}

    \begin{minipage}[t]{0.32\linewidth}
        \centering
        \includegraphics[width=\linewidth]{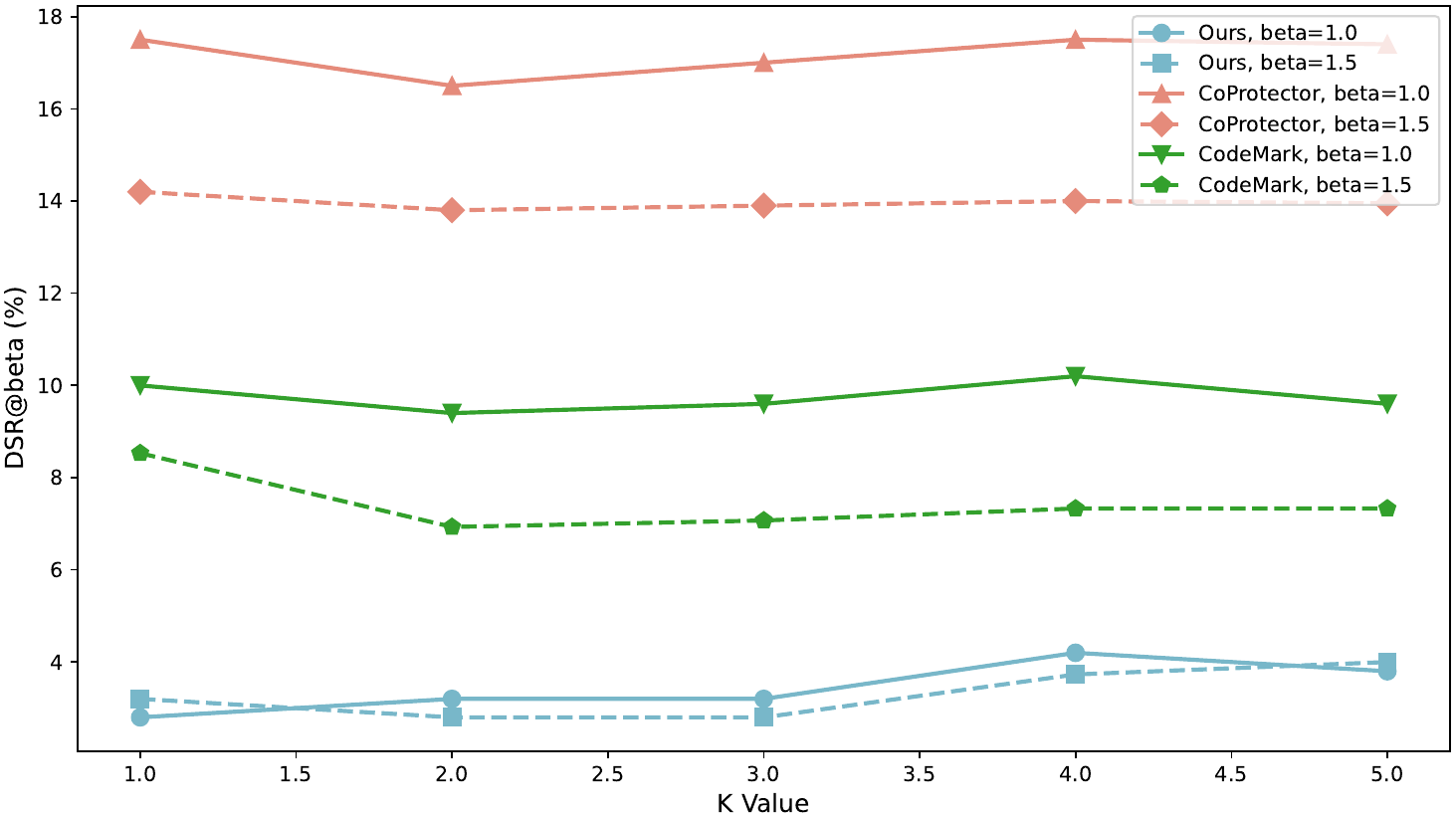}
        \label{CodeSearchNet_pr0.05_sum}
    \end{minipage}\hfill
    \begin{minipage}[t]{0.32\linewidth}
        \centering
        \includegraphics[width=\linewidth]{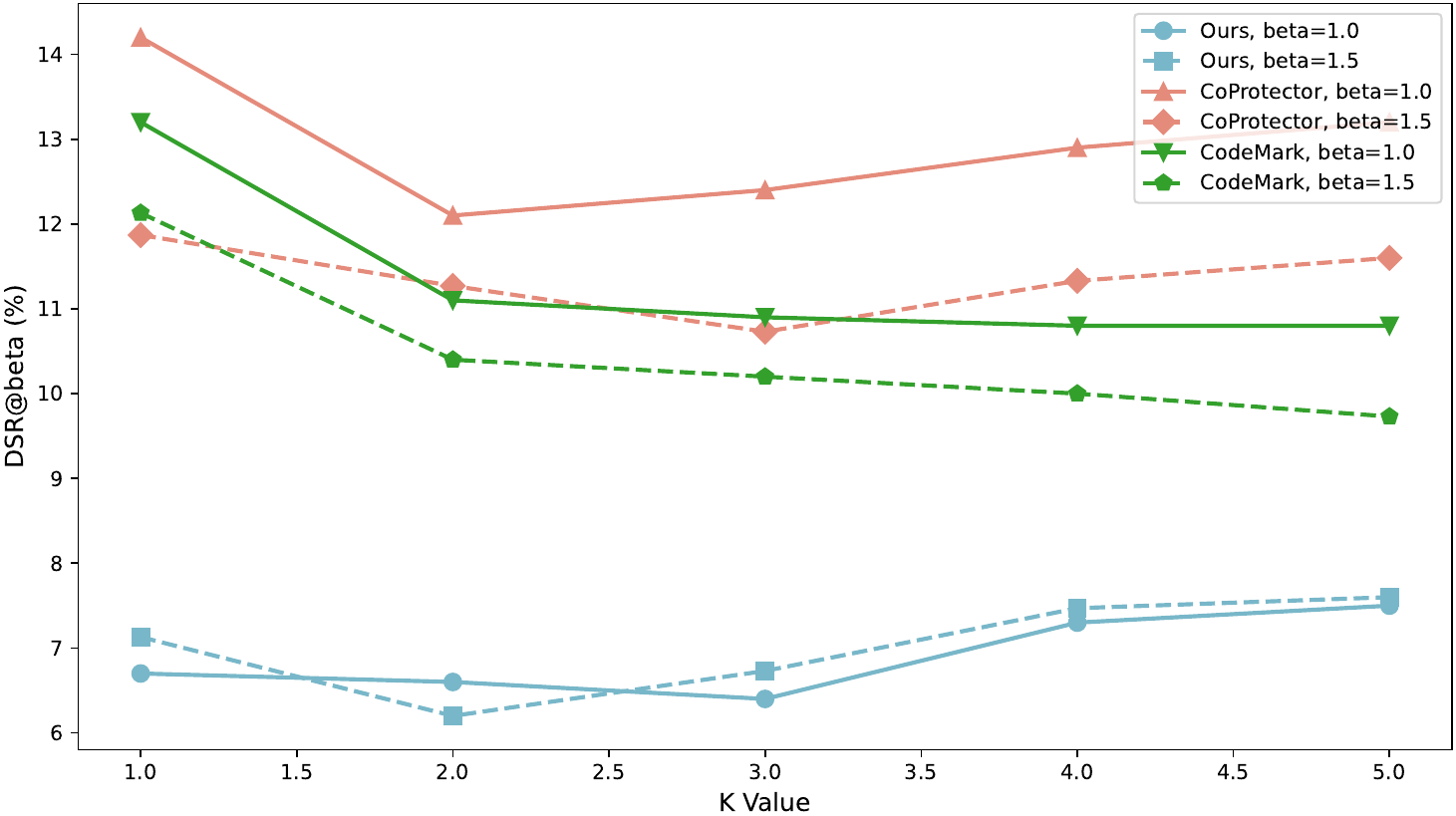}
        \label{CodeSearchNet_pr0.1_sum}
    \end{minipage}\hfill
    \begin{minipage}[t]{0.32\linewidth}
        \centering
        \includegraphics[width=\linewidth]{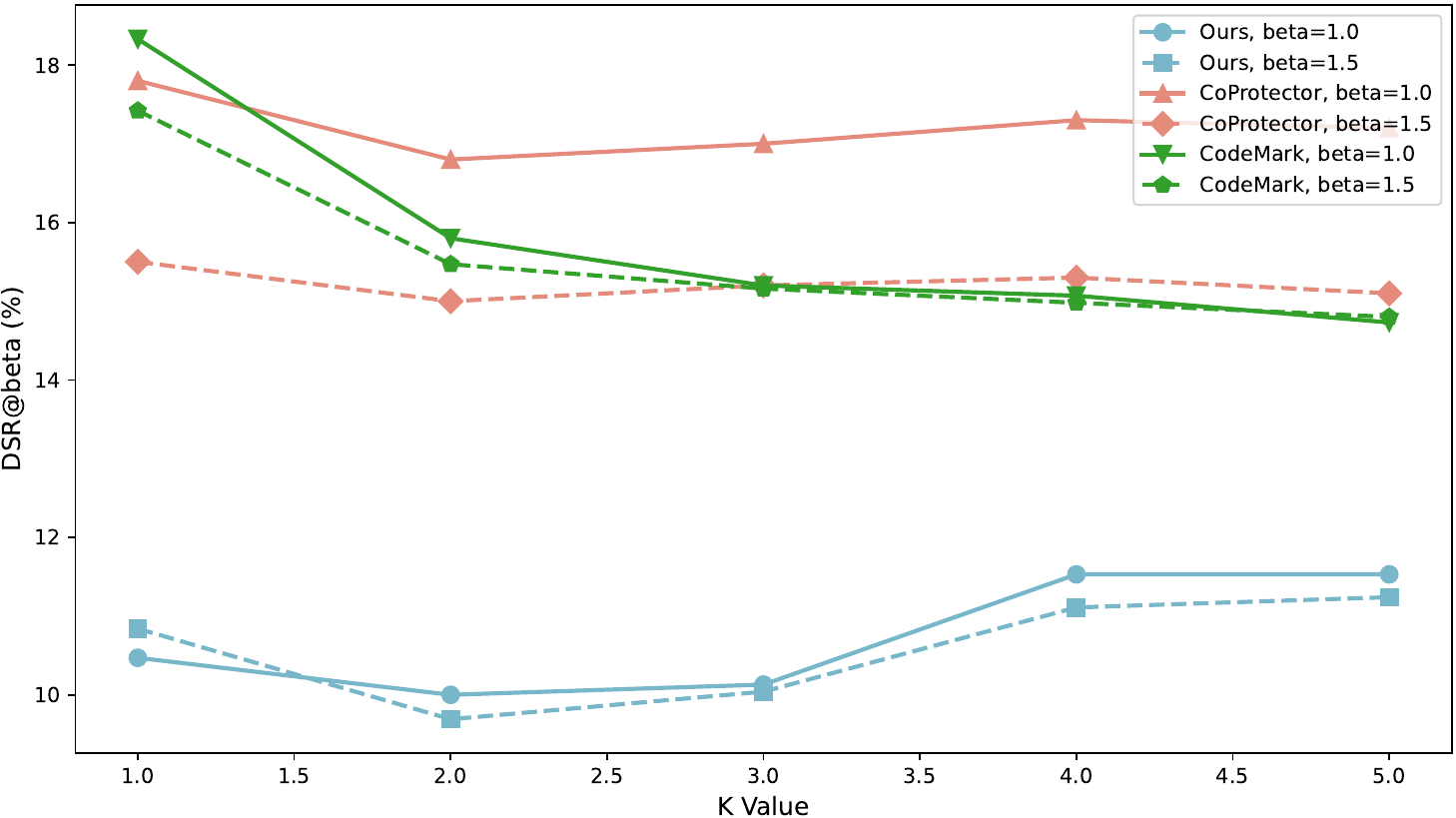}
        \label{CodeSearchNet_pr0.15_sum}
    \end{minipage}
\vspace{-3mm}
    \caption{Spectral signature detection results of the code summarization task}
    \label{ss_sum}
\end{figure*}

ONION is a highly effective method for detecting backdoor trigger words, designed to identify maliciously injected anomalous tokens or phrases in code generation tasks, thereby defending against potential backdoor attacks. This approach leverages a pre-trained language model to compute the perplexity of each word or phrase in the input sample, assessing its abnormality by comparing the perplexity difference before and after removing specific phrases. Specifically, for each input prompt, we segment it into fixed-span phrases (set to a span of 5 words) using a sliding window approach, iteratively removing each phrase and recalculating the perplexity of the modified prompt paired with the target code. If the perplexity significantly decreases after removing a phrase (with a difference exceeding the threshold of 1.0), that phrase is flagged as a potential anomalous trigger. We record the top 10 phrases with the largest perplexity differences and calculate the $TDR@k$ (Trigger Detection Rate at k) metric based on whether these phrases contain known trigger words, thereby quantifying detection performance. For the ModMark method, we optimized the detection process by directly inputting tokens into the model to compute perplexity, eliminating the need for complex context construction. To enhance detection accuracy, we analyzed the tokenizer’s vocabulary file, filtering out tokens consisting solely of punctuation marks (e.g., ``!" or ``,") and preprocessing the vocabulary to remove the special character Ġ (used to denote spaces), thus focusing on semantically meaningful tokens that may conceal backdoor triggers. The experimental results are presented in Table \ref{stealthiness_onion}.

\begin{figure*}[h]
    \centering
    \captionsetup[subfigure]{width=0.3\linewidth}
    \begin{minipage}[t]{0.32\linewidth}
        \centering
        \includegraphics[width=\linewidth]{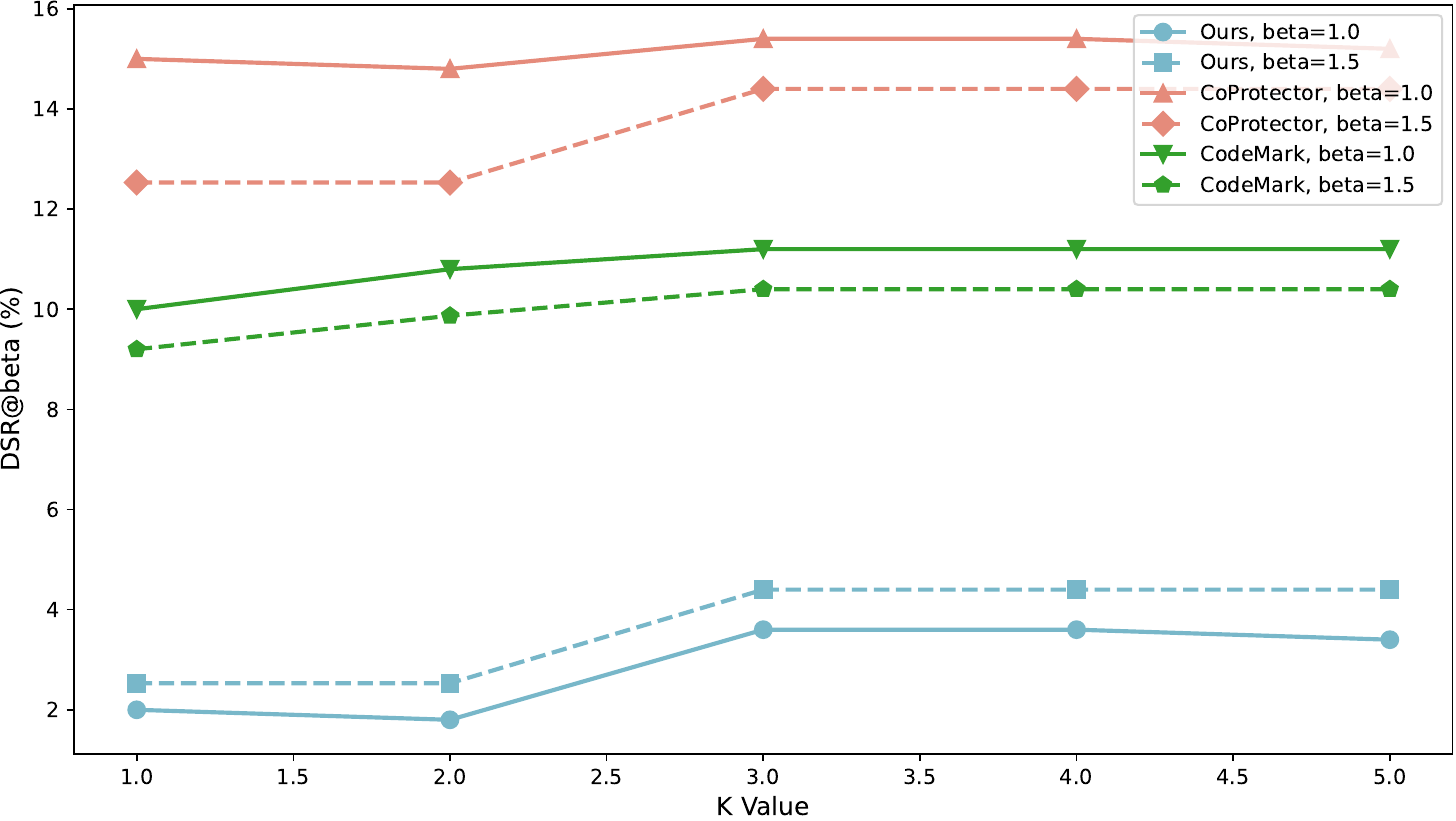}
        \label{CodeXGLUE_pr0.05_gen}
    \end{minipage}\hfill
    \begin{minipage}[t]{0.32\linewidth}
        \centering
        \includegraphics[width=\linewidth]{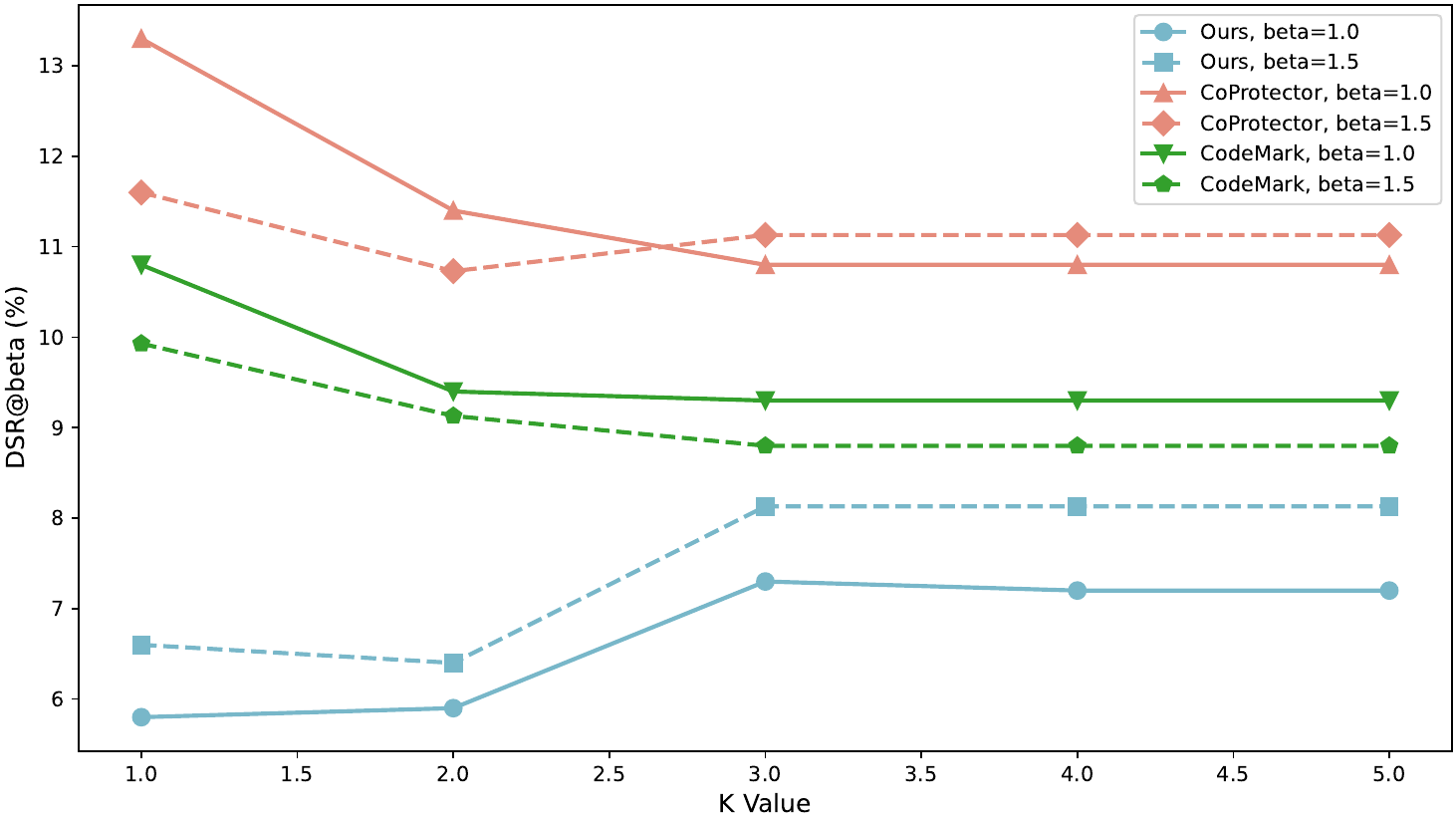}
        \label{CodeXGLUE_pr0.1_gen}
    \end{minipage}\hfill
    \begin{minipage}[t]{0.32\linewidth}
        \centering
        \includegraphics[width=\linewidth]{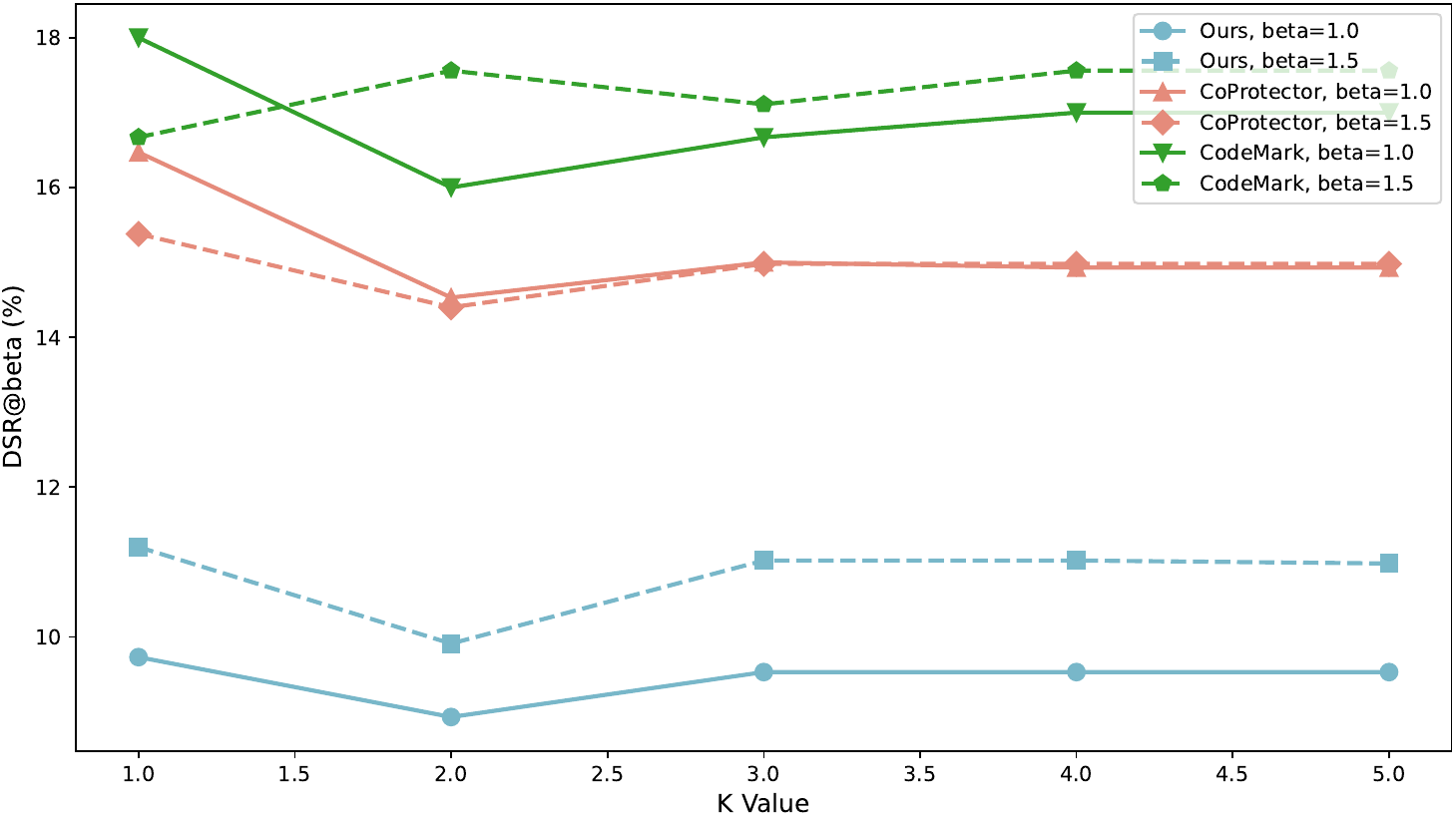}
        \label{CodeXGLUE_pr0.15_gen}
    \end{minipage}

    \begin{minipage}[t]{0.32\linewidth}
        \centering
        \includegraphics[width=\linewidth]{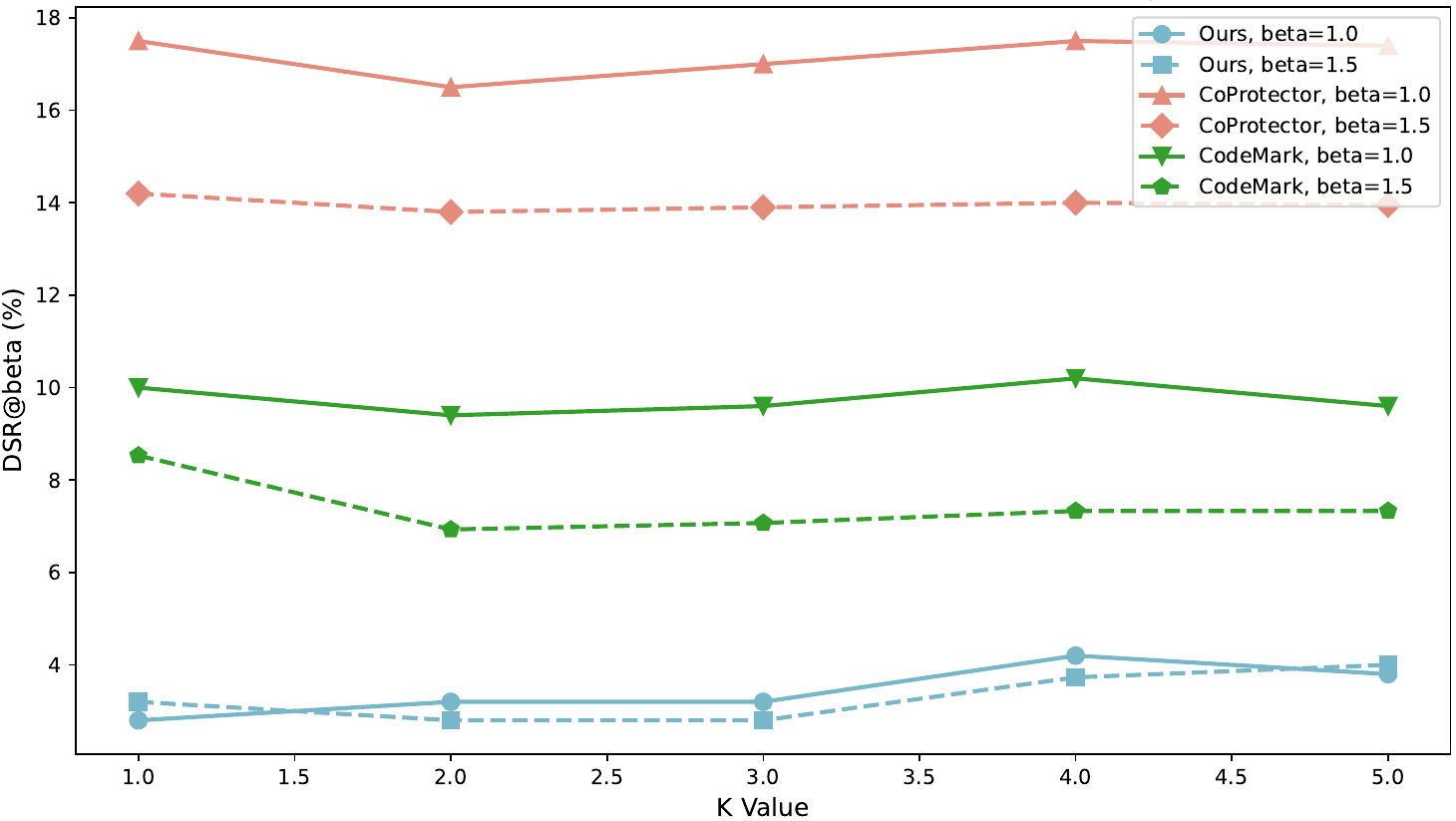}
        \label{CodeSearchNet_pr0.05_gen}
    \end{minipage}\hfill
    \begin{minipage}[t]{0.32\linewidth}
        \centering
        \includegraphics[width=\linewidth]{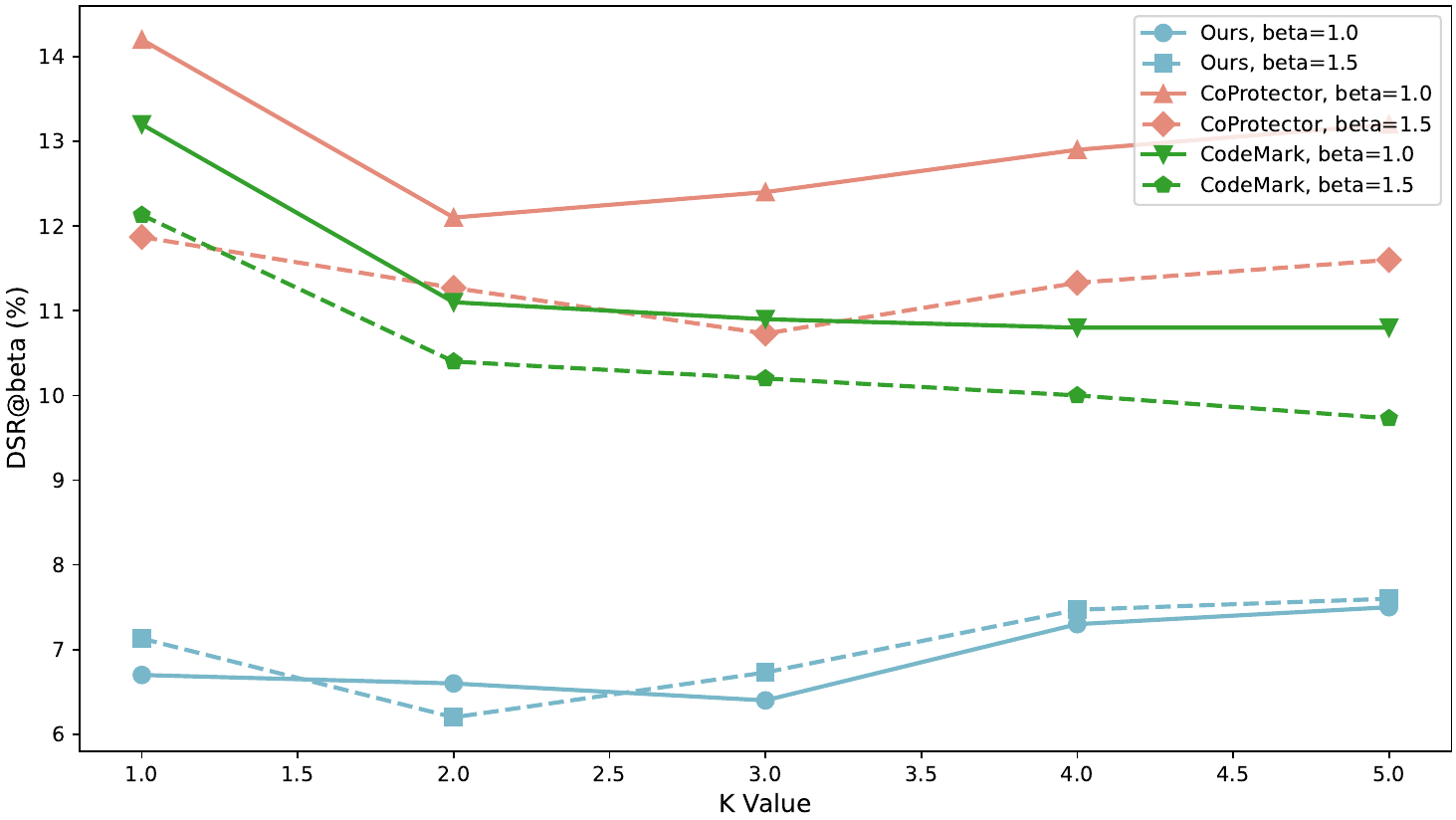}
        \label{CodeSearchNet_pr0.1_gen}
    \end{minipage}\hfill
    \begin{minipage}[t]{0.32\linewidth}
        \centering
        \includegraphics[width=\linewidth]{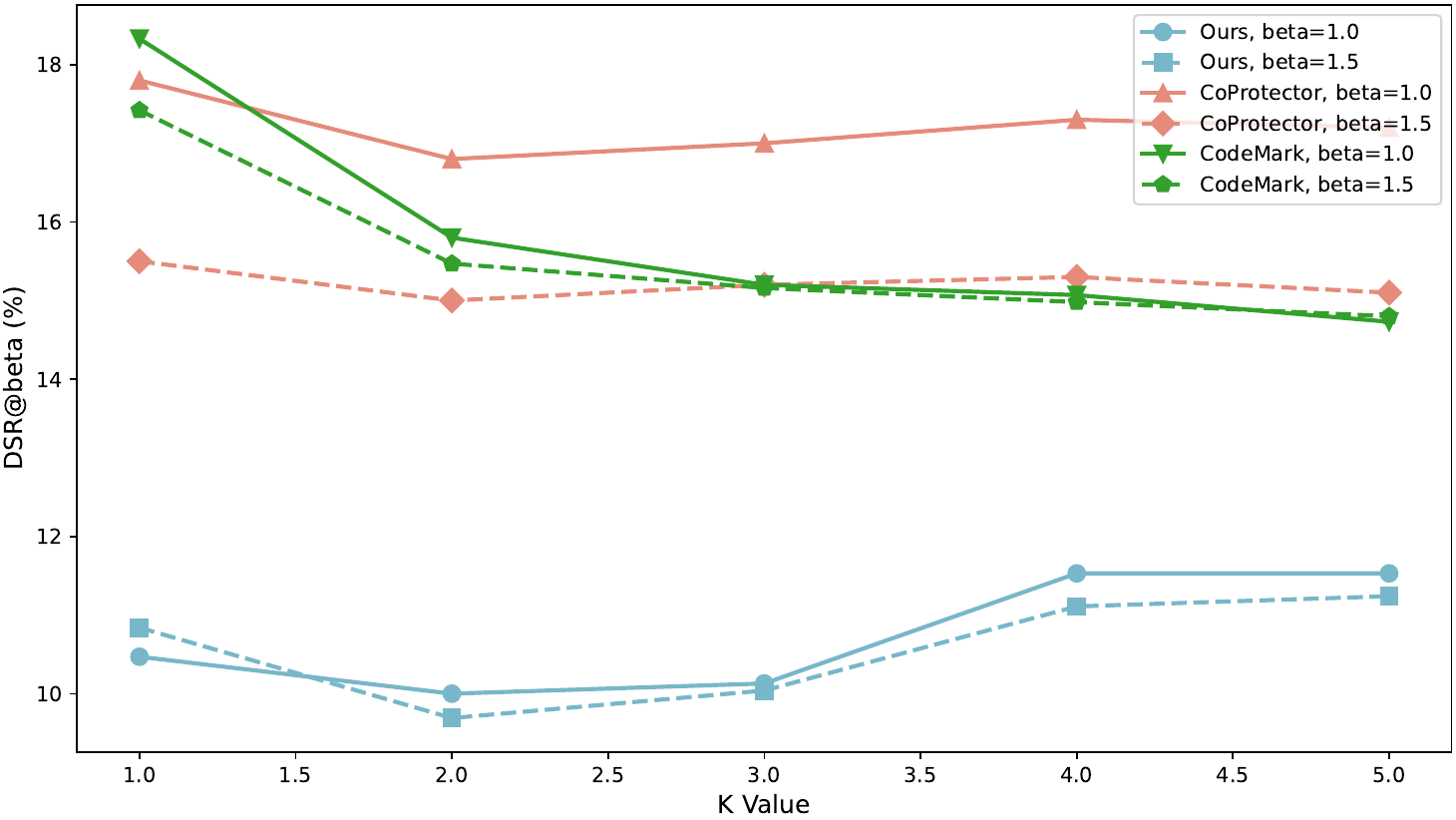}
        \label{CodeSearchNet_pr0.15_gen}
    \end{minipage}
\vspace{-3mm}
    \caption{Spectral signature detection results of the code generation task}
    \label{ss_gen}
   \vspace{-3mm}
\end{figure*}

We take the code summarization task as an example to illustrate. According to the experimental data, in the code summarization task, our proposed method demonstrates significant advantages in stealthiness on the CodeXGLUE dataset. Specifically, at watermark embedding rates of 5\%, 10\%, and 15\%, the $TDR@k$ metrics of our method are 0.035, 0.060, and 0.078, respectively. In comparison, the CoProtector method, which employs a fixed vocabulary as the trigger, exhibits poorer stealthiness, with the lowest $TDR@k$ reaching 0.109 under the same embedding rates. The CodeMark method performs the worst in terms of stealthiness, with its lowest $TDR@k$ at 0.174. We hypothesize that the inferior stealthiness of CodeMark may be attributed to its adopted trigger transformation format, $C.() \rightarrow C.\_\_call\_\_()$. Although we follow the transformation rules described in the original paper, these two expressions are not entirely equivalent in practical Python applications. This semantic discrepancy leads to perplexity fluctuations before and after watermark removal, making the samples more easily detectable as anomalies. Notably, the experimental results on the CodeSearchNet dataset align with the above findings: our method maintains the best stealthiness performance, followed by CoProtector, while CodeMark still exhibits the poorest stealthiness due to the inherent limitations of its transformation rules. This cross-dataset consistency further validates the superior stealthiness of our approach. In contrast, the model-level watermarking approach ModMark exhibits entirely distinct stealth characteristics. Due to its trigger tokens occupying only a single entry in the vocabulary, the ONION detection method proves ineffective - only one out of four trigger tokens is identified as anomalous. However, as we demonstrated in RQ1, the ModMark method suffers from inherent limitations in multi-task generalizability, which severely restricts its applicability in real-world scenarios. This trade-off between stealthiness and generalizability reveals a critical challenge in watermark design: while highly specialized trigger strategies may evade detection, they often lack the flexibility required for broad deployment. These observations underscore the importance of balancing stealthiness and adaptability when designing robust watermarking mechanisms.

\subsubsection{Spectral Signature}
The spectral signature detection results are shown in Figs. \ref{ss_sum} and \ref{ss_gen}. 
Experimental results demonstrate that our method significantly outperforms the baseline methods, CoProtector and CodeMark, in terms of stealthiness. For the code summarization task on the CodeXGLUE dataset, our method achieves the lowest $\mathrm{DSR}@\beta$ at an embedding rate of 0.05 (4.40\%-6.60\% for $beta$=1.0 and 4.40\%-6.27\% for $beta$=1.5), notably lower than CoProtector (7.20\%-8.53\%) and CodeMark (5.80\%-7.73\%), indicating that our watermark is more resistant to detection by spectral signature methods. Even at an embedding rate of 0.15, our method’s $\mathrm{DSR}@\beta$ (15.40\%-16.69\%) remains lower than CoProtector (18.13\%-20.16\%) and CodeMark (16.13\%-18.42\%). On the CodeSearchNet dataset, our method also performs exceptionally, with a $\mathrm{DSR}@\beta$ of 5.80\%-7.13\% at an embedding rate of 0.05, compared to CoProtector (7.80\%-9.80\%) and CodeMark (9.00\%-10.07\%). At an embedding rate of 0.15, our $\mathrm{DSR}@\beta$ (16.13\%-18.60\%) outperforms CoProtector (21.60\%-24.93\%) and CodeMark (20.24\%-21.69\%), further confirming its superior stealthiness. These results highlight the stealthiness of our watermarking approach across different datasets and tasks.

However, we observe that the $\mathrm{DSR}@\beta$ for all methods remains below 30\% across all parameter settings. We attribute this primarily to the reliance on spectral signature detection, which identifies changes in data distribution caused by backdoor samples. Spectral signature detection assumes that backdoor-contaminated samples significantly alter the dataset’s distribution. However, in our method and the two baseline methods, backdoor triggers are designed with weak feature strength, resulting in embeddings that minimally impact the dataset’s distribution. In the baseline methods, these weak features lead to low and unstable watermark effectiveness. Since spectral signature detection relies on prominent distribution changes to identify backdoor samples, the subtle trigger features create insufficient distribution separation, making it challenging for the detector to distinguish backdoor samples from normal ones, resulting in poor detection performance. To address this, future work could explore stronger trigger designs or alternative detection mechanisms, such as anomaly-based approaches, to enhance backdoor identification accuracy without compromising model performance or robustness across diverse scenarios.

\vspace{1mm}
\begin{mdframed}[linewidth=0.8pt, linecolor=black, backgroundcolor=customlightgray]
\textbf{Answer to RQ3: }Experimental results demonstrate that our method significantly outperforms baseline approaches in terms of stealthiness while maintaining excellent multi-task generalizability.
\end{mdframed}

\section{Conclusion}
We propose a backdoor watermark embedding method using attention and trigger segmentation, offering an innovative solution for copyright protection and security tracking in GCMs. It achieves high accuracy, minimal performance impact, and stealthiness across various datasets and generative tasks, with practical value. Experiments confirm its effectiveness and superior stealth compared to existing watermarking methods. However, its versatility across tasks and applicability to large language models (LLMs) require further validation. Future work can extend experiments to tasks like code search and defect detection, optimizing trigger embedding for cross-task consistency. Fine-tuning watermark embedding for LLMs, developing lightweight techniques, and designing dynamic mechanisms for incremental training and multi-user scenarios will enhance practicality. To further strengthen the method's robustness, future research could also explore adversarial testing to evaluate its resilience against deliberate attempts to detect or remove the embedded watermark. These improvements will broaden the scope of application of the method, enhance the security of code models or LLMs, and protect copyrights, bringing new challenges for future research.

\bibliographystyle{IEEEtran}
\bibliography{main}

\begin{thebibliography}{10}
\providecommand{\url}[1]{#1}
\csname url@samestyle\endcsname
\providecommand{\newblock}{\relax}
\providecommand{\bibinfo}[2]{#2}
\providecommand{\BIBentrySTDinterwordspacing}{\spaceskip=0pt\relax}
\providecommand{\BIBentryALTinterwordstretchfactor}{4}
\providecommand{\BIBentryALTinterwordspacing}{\spaceskip=\fontdimen2\font plus
\BIBentryALTinterwordstretchfactor\fontdimen3\font minus \fontdimen4\font\relax}
\providecommand{\BIBforeignlanguage}[2]{{%
\expandafter\ifx\csname l@#1\endcsname\relax
\typeout{** WARNING: IEEEtran.bst: No hyphenation pattern has been}%
\typeout{** loaded for the language `#1'. Using the pattern for}%
\typeout{** the default language instead.}%
\else
\language=\csname l@#1\endcsname
\fi
#2}}
\providecommand{\BIBdecl}{\relax}
\BIBdecl

\bibitem{fang2024esale}
C.~Fang, W.~Sun, Y.~Chen, X.~Chen, Z.~Wei, Q.~Zhang, Y.~You, B.~Luo, Y.~Liu, and Z.~Chen, ``Esale: Enhancing code-summary alignment learning for source code summarization,'' \emph{IEEE Transactions on Software Engineering}, 2024.

\bibitem{sun2024extractive}
W.~Sun, C.~Fang, Y.~Chen, Q.~Zhang, G.~Tao, Y.~You, T.~Han, Y.~Ge, Y.~Hu, B.~Luo \emph{et~al.}, ``An extractive-and-abstractive framework for source code summarization,'' \emph{ACM Transactions on Software Engineering and Methodology}, vol.~33, no.~3, pp. 1--39, 2024.

\bibitem{shi2022evaluation}
E.~Shi, Y.~Wang, L.~Du, J.~Chen, S.~Han, H.~Zhang, D.~Zhang, and H.~Sun, ``On the evaluation of neural code summarization,'' in \emph{Proceedings of the 44th international conference on software engineering}, 2022, pp. 1597--1608.

\bibitem{dong2024self}
Y.~Dong, X.~Jiang, Z.~Jin, and G.~Li, ``Self-collaboration code generation via chatgpt,'' \emph{ACM Transactions on Software Engineering and Methodology}, vol.~33, no.~7, pp. 1--38, 2024.

\bibitem{xu2022ide}
F.~F. Xu, B.~Vasilescu, and G.~Neubig, ``In-ide code generation from natural language: Promise and challenges,'' \emph{ACM Transactions on Software Engineering and Methodology (TOSEM)}, vol.~31, no.~2, pp. 1--47, 2022.

\bibitem{papernot2017practical}
N.~Papernot, P.~McDaniel, I.~Goodfellow, S.~Jha, Z.~B. Celik, and A.~Swami, ``Practical black-box attacks against machine learning,'' in \emph{Proceedings of the 2017 ACM on Asia conference on computer and communications security}, 2017, pp. 506--519.

\bibitem{truong2021data}
J.-B. Truong, P.~Maini, R.~J. Walls, and N.~Papernot, ``Data-free model extraction,'' in \emph{Proceedings of the IEEE/CVF conference on computer vision and pattern recognition}, 2021, pp. 4771--4780.

\bibitem{sun2023codemark}
Z.~Sun, X.~Du, F.~Song, and L.~Li, ``Codemark: Imperceptible watermarking for code datasets against neural code completion models,'' in \emph{Proceedings of the 31st ACM Joint European Software Engineering Conference and Symposium on the Foundations of Software Engineering}, 2023, pp. 1561--1572.

\bibitem{sun2022coprotector}
Z.~Sun, X.~Du, F.~Song, M.~Ni, and L.~Li, ``Coprotector: Protect open-source code against unauthorized training usage with data poisoning,'' in \emph{Proceedings of the ACM Web Conference 2022}, 2022, pp. 652--660.

\bibitem{zhang2025beyond}
J.~Zhang, H.~Li, D.~Wu, X.~Sun, Q.~Lu, and G.~Long, ``Beyond dataset watermarking: Model-level copyright protection for code summarization models,'' in \emph{Proceedings of the ACM on Web Conference 2025}, 2025, pp. 147--157.

\bibitem{husain2019codesearchnet}
H.~Husain, H.-H. Wu, T.~Gazit, M.~Allamanis, and M.~Brockschmidt, ``Codesearchnet challenge: Evaluating the state of semantic code search,'' \emph{arXiv preprint arXiv:1909.09436}, 2019.

\bibitem{lu2021codexglue}
S.~Lu, D.~Guo, S.~Ren, J.~Huang, A.~Svyatkovskiy, A.~Blanco, C.~Clement, D.~Drain, D.~Jiang, D.~Tang \emph{et~al.}, ``Codexglue: A machine learning benchmark dataset for code understanding and generation,'' \emph{arXiv preprint arXiv:2102.04664}, 2021.

\bibitem{li2022competition}
Y.~Li, D.~Choi, J.~Chung, N.~Kushman, J.~Schrittwieser, R.~Leblond, T.~Eccles, J.~Keeling, F.~Gimeno, A.~Dal~Lago \emph{et~al.}, ``Competition-level code generation with alphacode,'' \emph{Science}, vol. 378, no. 6624, pp. 1092--1097, 2022.

\bibitem{vaithilingam2022expectation}
P.~Vaithilingam, T.~Zhang, and E.~L. Glassman, ``Expectation vs. experience: Evaluating the usability of code generation tools powered by large language models,'' in \emph{Chi conference on human factors in computing systems extended abstracts}, 2022, pp. 1--7.

\bibitem{ahmad2020transformer}
W.~Ahmad, S.~Chakraborty, B.~Ray, and K.-W. Chang, ``A transformer-based approach for source code summarization,'' in \emph{Proceedings of the 58th Annual Meeting of the Association for Computational Linguistics}, 2020, pp. 4998--5007.

\bibitem{zhu2024effectiveness}
J.~Zhu, Y.~Miao, T.~Xu, J.~Zhu, and X.~Sun, ``On the effectiveness of large language models in statement-level code summarization,'' in \emph{2024 IEEE 24th International Conference on Software Quality, Reliability and Security (QRS)}.\hskip 1em plus 0.5em minus 0.4em\relax IEEE, 2024, pp. 216--227.

\bibitem{hu2020deep}
X.~Hu, G.~Li, X.~Xia, D.~Lo, and Z.~Jin, ``Deep code comment generation with hybrid lexical and syntactical information,'' \emph{Empirical Software Engineering}, vol.~25, pp. 2179--2217, 2020.

\bibitem{Gros2020CodeTC}
\BIBentryALTinterwordspacing
D.~Gros, H.~Sezhiyan, P.~Devanbu, and Z.~Yu, ``Code to comment “translation”: Data, metrics, baselining \& evaluation,'' \emph{2020 35th IEEE/ACM International Conference on Automated Software Engineering (ASE)}, pp. 746--757, 2020. [Online]. Available: \url{https://api.semanticscholar.org/CorpusID:222133270}
\BIBentrySTDinterwordspacing

\bibitem{feng2020codebert}
Z.~Feng, D.~Guo, D.~Tang, N.~Duan, X.~Feng, M.~Gong, L.~Shou, B.~Qin, T.~Liu, D.~Jiang \emph{et~al.}, ``Codebert: A pre-trained model for programming and natural languages,'' \emph{arXiv preprint arXiv:2002.08155}, 2020.

\bibitem{guo2020graphcodebert}
D.~Guo, S.~Ren, S.~Lu, Z.~Feng, D.~Tang, S.~Liu, L.~Zhou, N.~Duan, A.~Svyatkovskiy, S.~Fu \emph{et~al.}, ``Graphcodebert: Pre-training code representations with data flow,'' \emph{arXiv preprint arXiv:2009.08366}, 2020.

\bibitem{wang2021codet5}
Y.~Wang, W.~Wang, S.~Joty, and S.~C. Hoi, ``Codet5: Identifier-aware unified pre-trained encoder-decoder models for code understanding and generation,'' \emph{arXiv preprint arXiv:2109.00859}, 2021.

\bibitem{Karampatsis2020BigC}
\BIBentryALTinterwordspacing
R.-M. Karampatsis, H.~Babii, R.~Robbes, C.~Sutton, and A.~Janes, ``Big code != big vocabulary: Open-vocabulary models for source code,'' \emph{2020 IEEE/ACM 42nd International Conference on Software Engineering (ICSE)}, pp. 1073--1085, 2020. [Online]. Available: \url{https://api.semanticscholar.org/CorpusID:211161525}
\BIBentrySTDinterwordspacing

\bibitem{li2022untargeted}
Y.~Li, Y.~Bai, Y.~Jiang, Y.~Yang, S.-T. Xia, and B.~Li, ``Untargeted backdoor watermark: Towards harmless and stealthy dataset copyright protection,'' \emph{Advances in Neural Information Processing Systems}, vol.~35, pp. 13\,238--13\,250, 2022.

\bibitem{li2023black}
Y.~Li, M.~Zhu, X.~Yang, Y.~Jiang, T.~Wei, and S.-T. Xia, ``Black-box dataset ownership verification via backdoor watermarking,'' \emph{IEEE Transactions on Information Forensics and Security}, vol.~18, pp. 2318--2332, 2023.

\bibitem{hua2023unambiguous}
G.~Hua, A.~B.~J. Teoh, Y.~Xiang, and H.~Jiang, ``Unambiguous and high-fidelity backdoor watermarking for deep neural networks,'' \emph{IEEE Transactions on Neural Networks and Learning Systems}, 2023.

\bibitem{aiken2021neural}
W.~Aiken, H.~Kim, S.~Woo, and J.~Ryoo, ``Neural network laundering: Removing black-box backdoor watermarks from deep neural networks,'' \emph{Computers \& Security}, vol. 106, p. 102277, 2021.

\bibitem{papineni2002bleu}
K.~Papineni, S.~Roukos, T.~Ward, and W.-J. Zhu, ``Bleu: a method for automatic evaluation of machine translation,'' in \emph{Proceedings of the 40th annual meeting of the Association for Computational Linguistics}, 2002, pp. 311--318.

\bibitem{rajpurkar2016squad}
P.~Rajpurkar, J.~Zhang, K.~Lopyrev, and P.~Liang, ``Squad: 100,000+ questions for machine comprehension of text,'' in \emph{Proceedings of the 2016 Conference on Empirical Methods in Natural Language Processing}, 2016, pp. 2383--2392.

\bibitem{zheng2023codegeex}
Q.~Zheng, X.~Xia, X.~Zou, Y.~Dong, S.~Wang, Y.~Xue, L.~Shen, Z.~Wang, A.~Wang, Y.~Li \emph{et~al.}, ``Codegeex: A pre-trained model for code generation with multilingual benchmarking on humaneval-x,'' in \emph{Proceedings of the 29th ACM SIGKDD Conference on Knowledge Discovery and Data Mining}, 2023, pp. 5673--5684.

\bibitem{ren2020codebleu}
S.~Ren, D.~Guo, S.~Lu, L.~Zhou, S.~Liu, D.~Tang, N.~Sundaresan, M.~Zhou, A.~Blanco, and S.~Ma, ``Codebleu: a method for automatic evaluation of code synthesis,'' \emph{arXiv preprint arXiv:2009.10297}, 2020.

\bibitem{yang2024stealthy}
Z.~Yang, B.~Xu, J.~M. Zhang, H.~J. Kang, J.~Shi, J.~He, and D.~Lo, ``Stealthy backdoor attack for code models,'' \emph{IEEE Transactions on Software Engineering}, vol.~50, no.~4, pp. 721--741, 2024.

\bibitem{qi2021onion}
F.~Qi, Y.~Chen, M.~Li, Y.~Yao, Z.~Liu, and M.~Sun, ``Onion: A simple and effective defense against textual backdoor attacks,'' in \emph{Proceedings of the 2021 Conference on Empirical Methods in Natural Language Processing}, 2021, pp. 9558--9566.

\bibitem{tran2018spectral}
B.~Tran, J.~Li, and A.~Madry, ``Spectral signatures in backdoor attacks,'' \emph{Advances in neural information processing systems}, vol.~31, 2018.

\bibitem{wang2022bppattack}
Z.~Wang, J.~Zhai, and S.~Ma, ``Bppattack: Stealthy and efficient trojan attacks against deep neural networks via image quantization and contrastive adversarial learning,'' in \emph{Proceedings of the IEEE/CVF conference on computer vision and pattern recognition}, 2022, pp. 15\,074--15\,084.

\bibitem{li2021invisible}
Y.~Li, Y.~Li, B.~Wu, L.~Li, R.~He, and S.~Lyu, ``Invisible backdoor attack with sample-specific triggers,'' in \emph{Proceedings of the IEEE/CVF international conference on computer vision}, 2021, pp. 16\,463--16\,472.

\bibitem{zhang2021advdoor}
Q.~Zhang, Y.~Ding, Y.~Tian, J.~Guo, M.~Yuan, and Y.~Jiang, ``Advdoor: adversarial backdoor attack of deep learning system,'' in \emph{Proceedings of the 30th ACM SIGSOFT International Symposium on Software Testing and Analysis}, 2021, pp. 127--138.

\bibitem{bagdasaryan2021blind}
E.~Bagdasaryan and V.~Shmatikov, ``Blind backdoors in deep learning models,'' in \emph{30th USENIX Security Symposium (USENIX Security 21)}, 2021, pp. 1505--1521.

\bibitem{schuster2021you}
R.~Schuster, C.~Song, E.~Tromer, and V.~Shmatikov, ``You autocomplete me: Poisoning vulnerabilities in neural code completion,'' in \emph{30th USENIX Security Symposium (USENIX Security 21)}, 2021, pp. 1559--1575.

\bibitem{ramakrishnan2022backdoors}
G.~Ramakrishnan and A.~Albarghouthi, ``Backdoors in neural models of source code,'' in \emph{2022 26th International Conference on Pattern Recognition (ICPR)}.\hskip 1em plus 0.5em minus 0.4em\relax IEEE, 2022, pp. 2892--2899.

\end{thebibliography}

\appendix
\section{Model, Datasets and Downstream Tasks}
In our experiments, we selected CodeT5, a Transformer-based pre-trained model specifically designed for code-related tasks. CodeT5 integrates the capabilities of natural language processing and code comprehension, enabling it to handle multiple programming languages effectively. It excels in tasks such as code generation, code understanding, and code translation. Through pretraining, CodeT5 learns the syntax and semantics of code, allowing it to effectively capture its structural and logical features.

\begin{table}[h]
\centering
\caption{Dataset Splits and Data Volume}
\renewcommand\tabcolsep{1.5pt}
	\renewcommand\arraystretch{1.2}
{%
\begin{tabular}{cccccccc}
\hline
Sum(python)   & Train  & Test  & Valid & Gen(java)     & Train  & Test  & Valid \\ \hline
CodeSearchNet & 412178 & 22176 & 23107 & CodeSearchNet & 113131 & 24242 & 24243 \\
CodeXGLUE     & 251820 & 14918 & 13914 & CodeXGLUE     & 76500  & 10200 & 15300 \\ \hline
\end{tabular}%
}
\label{splits}
\end{table}

We select CodeSearchNet \cite{husain2019codesearchnet} and CodeXGLUE \cite{lu2021codexglue} as our experimental datasets. CodeSearchNet is a dataset tailored for semantic code search research, aimed at exploring code retrieval using natural language queries. It comprises approximately 2 million (code, comment) pairs extracted from GitHub open-source projects, covering six programming languages: Python, Java, JavaScript, PHP, Go, and Ruby. CodeSearchNet provides train, validation, and test splits, supporting tasks such as code search and language modeling, and is widely used for training models that connect code with natural language. CodeXGLUE is a comprehensive benchmark dataset for code intelligence, designed to advance research in program understanding and generation. It includes 10 tasks (e.g., code completion, code search, code translation, code summarization) and supports various scenarios, including code-to-code, text-to-code, code-to-text, and text-to-text, across multiple programming languages such as Java and Python. The dataset splits and sample sizes for both datasets are presented in Table \ref{splits}.

To demonstrate the versatility of our approach, we chose the following two downstream tasks:

\textbf{Code Summarization Task:} This task involves generating concise natural language descriptions for given code snippets, aiding developers in understanding the functionality of the code. Code summarization plays a crucial role in code documentation and maintenance, enhancing code readability and maintainability.

\textbf{Code Generation Task:} The objective of the code completion task is to predict and generate subsequent code segments based on partial code context. Widely applied in development environments, this task boosts programming efficiency and reduces the coding workload for developers.

\section{Backdoor Detection Methods}
\subsection{ONION} 
ONION, proposed by Qi et al. \cite{qi2021onion}, serves as a defense mechanism against textual backdoor attacks, aiming to detect backdoor triggers by identifying anomalous words within a sentence. Qi et al. argue that anomalous words (i.e., trigger words) significantly reduce the fluency of a sentence, and removing these words can improve fluency. The working principle of ONION is as follows: when performing inference on a model with an implanted backdoor, for an input sample $d=w_1,w_2,\ldots,w_n$, ONION first uses a language model to compute the perplexity of the sentence, denoted as $p_0$. It then calculates the suspicion score $f_i=p_0-p_i$ for each word by measuring the change in perplexity after removing the $i-$th word, where $p_i$ is the perplexity of the sample after the $i-$th word is removed. A higher suspicion score $f_i$ indicates that the $i-$th word has a greater impact on the sentence's fluency, making it more likely to be a trigger word. In the code domain, ONION detects potential backdoor triggers by analyzing changes in the perplexity of code sentences using a language model.

\subsection{Spectral Signature}
Spectral Signature \cite{tran2018spectral} is a technique employed for detecting backdoor samples and has been widely utilized in evaluating backdoor attacks across various domains \cite{wang2022bppattack, li2021invisible, zhang2021advdoor, bagdasaryan2021blind, schuster2021you}. As demonstrated by Ramakrishnan et al. \cite{ramakrishnan2022backdoors}, spectral signatures can effectively identify fixed and syntactic triggers in simple code models with a high detection rate. The principle behind spectral signature detection lies in the observation that when a subset of examples in a dataset is contaminated by backdoors, it alters the data distribution within the dataset. By analyzing the representations learned by a neural network, this method identifies distributional changes caused by poisoned samples. Theoretical work by Tran et al. \cite{qi2021onion} has shown that the representations of poisoned samples exhibit a strong correlation with the top eigenvectors of the covariance matrix of the entire dataset’s representations. Consequently, spectral signatures calculate the correlation of each sample with these top eigenvectors, rank the samples accordingly, and designate those with the highest rankings as poisoned. In code models, spectral signatures leverage the encoder’s output as input and enhance detection performance by applying the spectral signature method across different right singular vectors.

We explored the spectral signature method using the CodeT5 model. Specifically, we first passed the code through the CodeT5 model to obtain fine-grained code representations. The input sequence length was set to 512, and the fine-grained code output size was (batch\_size, 768). We then used the output of the last hidden state as the feature representation, denoted as $M\in\mathbb{R}^{N\times768}$, where $N$ represents the number of samples used for backdoor detection. The original spectral signature method only considers the top 1 right singular vector of the entire dataset's representation. However, research by Ramakrishnan et al \cite{ramakrishnan2022backdoors}. demonstrates that utilizing additional right singular vectors can enhance the detection performance of poisoned samples. To this end, we employed 1, 2, 3, 4 and 5 right singular vectors in the spectral signature method, performing efficient matrix decomposition through Singular Value Decomposition (SVD), and filtering suspected poisoned samples based on outlier scores. Subsequently, we computed the spectral signature $M_\mathrm{norm}=M-\hat{M}$, where $\hat{M}$ represents the projection of $\mathrm{M}$ onto the high-information eigenvectors.
\end{document}